\documentclass[journal]{IEEEtran}
\usepackage{amsmath,amssymb,amsfonts,amsthm,mathtools}
\usepackage{algorithm}
\usepackage{algorithmic}
\usepackage{array}
\usepackage{graphicx}
\usepackage{cite}
\usepackage{booktabs}
\usepackage{url}
\usepackage{xcolor}
\usepackage{multirow}
\usepackage{tabularx}
\usepackage{dsfont}
\usepackage[caption=false,font=normalsize,labelfont=sf,textfont=sf]{subfig}

\newtheorem{theorem}{Theorem}
\newtheorem{lemma}[theorem]{Lemma}
\newtheorem{corollary}[theorem]{Corollary}
\newtheorem{definition}[theorem]{Definition}
\newtheorem{remark}[theorem]{Remark}
\newtheorem{assumption}[theorem]{Assumption}

\begin{document}

\title{Degeneracy-Aware Resilient Resource Allocation in Cell-Free Cache-Aided MU-MIMO Networks}

\author{Sayanti~Ghosh,~\IEEEmembership{Member,~IEEE}, Indrakshi~Dey,~\IEEEmembership{Senior~Member,~IEEE}, and Nicola~Marchetti,~\IEEEmembership{Senior~Member,~IEEE}%
\thanks{S.~Ghosh and N.~Marchetti are with the Department of Electrical and Electronic Engineering, Trinity College Dublin, Dublin, Ireland (e-mail: saghosh@tcd.ie; nicola.marchetti@tcd.ie). I.~Dey is with the Walton Institute, South East Technological University, Waterford, Ireland (e-mail: indrakshi.dey@waltoninstitute.ie).}%
\thanks{Supported in part by the EU MSCA Project COALESCE under Grant 101130739, by US--Ireland R\&D Partnership RI-SFI-23/US/3924, and by Research Ireland Grant 13/RC/2077\_P2.}%
}

\maketitle

\begin{abstract}
Cell-free cache-aided multi-user multiple-input-multiple-output (MIMO) (CF-CA-MU-MIMO) networks improve spectral efficiency through coded multicast delivery and distributed spatial multiplexing, but their distributed architecture introduces vulnerabilities to jamming, cache-aware eavesdropping, Byzantine corruption, and pilot-contamination attacks. This paper develops a degeneracy-aware resilient framework based on four vulnerability-mode partitions (subfile, edge node, multicast stream, and user) and three attack-aware structural metrics: Degeneracy-Weighted Path Robustness (DWPR$^{\mathrm{att}}$), trust-aware Functional Substitution Score (FSS$^{\mathrm{trust}}$), and a robust degeneracy index ($D_k^{\mathrm{rob}}$). These metrics are incorporated into a fully decentralized consensus-based agent framework (DC-ABM) using trust-weighted trimmed-mean aggregation and adaptive trust evolution. Five theoretical results are established: (i) a tight top-mass concentration lemma, (ii) matching memory--rate--resilience achievability and converse bounds, (iii) a robust-degeneracy bound with outage characterization, (iv) a secrecy--cache coupling theorem, and (v) a Byzantine-robust mean-square convergence result with an explicit breakdown threshold $f_{\max}$. Simulations validate the analytical bounds and demonstrate $1.8\times$ to $3\times$ faster convergence than distributed alternating direction method of multipliers (ADMM), multi-agent reinforcement learning (MARL)/graph neural network (GNN)-based control, and Su--Vaidya consensus while maintaining throughput up to the predicted threshold $f_{\max}\approx0.19$.
\end{abstract}

\section{Introduction}
\label{sec:introduction}

Future IMT-2030 and 6G networks are expected to support massive connectivity, distributed intelligence, resilient communication, and scalable decentralized operation~\cite{ITU_IMT2030,3GPPRel19,ETSI_MAT_2026}. Among the candidate enabling technologies, cell-free massive multiple-input-multiple-output (MIMO) and coded caching have attracted significant attention because of their ability to improve spectral efficiency, macro-diversity, and content-delivery efficiency through cooperative transmission and multicast communication~\cite{Bjornson_CF,CellFree6G,MN_Caching,TolliLampiris}. Existing studies have investigated energy-efficient cell-free architectures, pilot-assisted transmission, multicast beamforming, and cache-aided MIMO precoding~\cite{Ngo_CellFree,Interdonato_CF,TolliLampiris,CacheBeamforming}. Edge-assisted caching has further been recognized as a key mechanism for reducing delivery latency and fronthaul congestion in future wireless systems~\cite{EdgeCaching6G}.

Cell-free cache-aided multi-user MIMO (CF-CA-MU-MIMO) systems jointly exploit distributed transmission and coded multicast delivery to improve spectral efficiency and content-delivery performance. Nevertheless, their distributed operation exposes the network to several security and robustness challenges, including pilot contamination, jamming, cache-aware eavesdropping, Byzantine corruption, and coordination failures, whose impact can propagate across multiple users through shared multicast transmissions~\cite{CellFreeSecurity,PilotAttack_RSMA}. In particular, multicast-coded transmissions couple multiple users through shared streams, allowing attacks affecting a single multicast stream to propagate across several users. Consequently, ensuring robustness becomes increasingly challenging as network size and coordination complexity grow.

To address scalability challenges, distributed optimization and learning-based approaches have been proposed for wireless resource allocation and decentralized coordination. Representative examples include decentralized optimization based on alternating direction method of multipliers (ADMM)~\cite{ADMMWireless}, reinforcement-learning-based resource allocation~\cite{MARL_Wireless}, graph-neural-network-assisted wireless control~\cite{GNN_Wireless}, and Byzantine-resilient distributed optimization and learning~\cite{SuVaidya,ByzantineLearning}. Although these methods improve scalability and robustness, they generally treat throughput optimization, caching, and adversarial resilience as separate design problems and do not explicitly capture the interaction among multicast coupling, cache structure, adversarial attacks, and decentralized fronthaul coordination.

Motivated by this gap, we develop a degeneracy-aware resilient coordination framework for CF-CA-MU-MIMO systems under concurrent jamming, cache-aware eavesdropping, and Byzantine attacks. The proposed framework builds upon the concepts of degeneracy and structural robustness~\cite{Giulio1999,dey2025degeneracy} and introduces three attack-aware structural metrics: Degeneracy-Weighted Path Robustness (DWPR$^{\mathrm{att}}$), trust-aware Functional Substitution Score (FSS$^{\mathrm{trust}}$), and a robust degeneracy index ($D_k^{\mathrm{rob}}$). These metrics are integrated into a fully decentralized consensus-based agent framework (DC-ABM) employing trust-weighted trimmed aggregation and local coordination without a central controller. The resulting framework jointly captures multicast robustness, structural substitutability, and adversarial feasibility degradation. Numerical results demonstrate significant gains in throughput robustness, convergence speed, scalability, and adversarial resilience compared with distributed ADMM, multi-agent reinforcement learning (MARL)/Graph Neural Network (GNN)-based control, Byzantine-resilient consensus, and centralized baselines.

\subsection{Main Contributions}
\label{subsec:contributions}

The main contributions of this paper are summarized as follows:

\begin{itemize}

\item We develop a degeneracy-aware resilient framework for CF-CA-MU-MIMO systems under concurrent jamming, pilot contamination, cache-aware eavesdropping, and Byzantine corruption.

\item We introduce attack-aware structural metrics, namely DWPR$^{\mathrm{att}}$, FSS$^{\mathrm{trust}}$, and the robust degeneracy index, to jointly characterize multicast robustness, structural substitutability, and adversarial feasibility degradation.

\item We propose a DC-ABM using trust-weighted trimmed aggregation and local distributed coordination without requiring centralized optimization.

\item We derive analytical bounds relating adversarial budgets, convergence stability, fronthaul spectral connectivity, caching diversity, and robust throughput.

\item Numerical results show that the proposed DC-ABM significantly improves throughput robustness, convergence speed, scalability, and adversarial resilience compared with distributed ADMM, MARL/GNN-based optimization, Byzantine-resilient consensus, and centralized baselines.

\end{itemize}

\begin{table*}[t]
\centering
\caption{Acronyms used in this paper.}
\label{tab:acronyms}
\renewcommand{\arraystretch}{1.0}
\footnotesize
\begin{tabular}{l l | l l}
\hline
\textbf{Acronym} & \textbf{Expansion} & \textbf{Acronym} & \textbf{Expansion} \\
\hline
6G, IMT-2030      & sixth-generation wireless / Int.\ Mobile Telecom.\ 2030 & LS              & least squares (channel estimator)\\
ABM / DC-ABM      & agent-based model / decentralized-consensus ABM         & MARL            & multi-agent reinforcement learning \\
ADMM              & alternating direction method of multipliers             & MN              & Maddah-Ali--Niesen (coded caching) \\
CF-CA-MU-MIMO     & cell-free cache-aided multi-user MIMO                   & MIMO / MU-MIMO  & multi-input multi-output / multi-user MIMO\\
CSI               & channel state information                               & MRR             & memory--rate--resilience (trade-off region)\\
D2D               & device-to-device (direct user link)                     & OMA             & orthogonal multiple access \\
DoF               & degrees of freedom                                      & RSMA            & rate-splitting multiple access \\
DWPR              & degeneracy-weighted path robustness                     & SDMA            & space-division multiple access \\
EN                & edge node (cell-free distributed antenna unit)          & SINR            & signal-to-interference-plus-noise ratio \\
ETSI / ISG-MAT    & European Telecom.\ Standards Inst.\ / MA study group    & TDD             & time-division duplex \\
FPR               & false-positive rate                                     & XOR             & exclusive-or (bitwise modulo-2 addition) \\
FSS               & functional substitution score                           & --              & --\\
GNN               & graph neural network                                    & --              & --\\
\hline
\end{tabular}
\end{table*}

\begin{table*}[t]
\centering
\caption{Principal symbols.}
\label{tab:symbols}
\renewcommand{\arraystretch}{1.0}
\footnotesize
\begin{tabular}{l l | l l}
\hline
\textbf{Symbol} & \textbf{Meaning} & \textbf{Symbol} & \textbf{Meaning} \\
\hline
$L,\,N_a,\,M_t\!=\!LN_a$            & no.\ ENs, antennas/EN, total spatial DoF      & $\mathcal{P}_{\mathrm{sub/EN/mcast/user}}$            & vulnerability-mode partitions \\
$K,\,G,\,K_g\!=\!K/G$               & total users, groups, users/group              & $\chi_{\mathcal{P}}(\mathcal{S},m),\,\boldsymbol{\alpha}^{\mathcal{P}}_k$ & mode incidence; rate-weighted mode-mass \\
$\mathcal{G}_f,\,\mathcal{G}_u,\,\lambda_2(\mathbf{L}_f)$ & fronthaul/D2D graphs; fronthaul spectral gap     & $\mathcal{J},\,\mathcal{E},\,\mathcal{B}$            & jammer, eavesdropper, Byzantine sets \\
$N,\,F$                             & library size (files); file size (bits)        & $b_{\mathcal{J}},\,b_{\mathcal{E}},\,b_{\mathcal{B}},\,f_{\mathcal{B}}$ & per-class budgets; Byzantine fraction \\
$M,\,t\!=\!K_g M/N$                 & per-user cache; MN cache fraction             & $P_{\mathrm{jam}}$                                       & per-stream jamming power \\
$\tau_c,\,\tau_p,\,\tau_d$          & coherence, pilot, data lengths (symbols)      & $\mathrm{DWPR}_k^{\mathrm{att}},\,\mathrm{FSS}_k^{\mathrm{trust}},\,D_k^{\mathrm{rob}}$ & three structural metrics \\
$\mathbf{h}_{k,\ell},\,\mathbf{h}_k$ & per-EN and aggregated channel                 & $\Psi_k^{\mathrm{adv}},\,\hat{\Psi}_k^{\mathrm{adv}}$    & composite resilience score; surrogate \\
$\beta_{k,\ell},\,\mathbf{R}_{k,\ell}$ & path loss; spatial correlation               & $\omega_i,\,\zeta_{\mathcal{P}}$                         & top-level / partition-salience weights \\
$\sigma_n^2,\,\sigma_{e,k}^2$       & noise; CSI estimation-error variance          & $\tau_j(t),\,\mu_j(t),\,\eta_\tau$                       & trust score, consistency, trust rate \\
$\mathbf{w}_{\mathcal{S}},\,\mathbf{w}_{\mathcal{S},\ell}$ & global / per-EN precoder for stream $\mathcal{S}$ & $\beta,\,\xi_{\mathrm{EN}},\,\xi_{\mathrm{user}}$ & trim parameter; consistency tolerances \\
$P_{\mathcal{S}},\,P_\ell^{\max}$   & stream power; per-EN power budget             & $\eta_t,\,\zeta$                                         & primal-dual step; dual mixing coeff. \\
$\gamma_{k,\mathcal{S}},\,R_{\mathcal{S}},\,R_k$ & per-stream SINR / per-stream / user rate          & $\Phi_k(\mathbf{b}),\,L_{\mathcal{J}},\,L_{\mathcal{B}}$ & resilience factor; jamming/Byz.\ loss \\
$\gamma_k^{\mathrm{tar}},\,D_k=\gamma_k^{\mathrm{tar}}/\gamma_k^p$
& target SINR; nominal degeneracy index & $\Delta^{\mathcal{J}}_k,\,\Delta^{\mathcal{E}}_k,\,\Delta^{\mathcal{B}}_k$ & class-specific $D_k^{\mathrm{rob}}$ perturbations \\
$c_{\mathcal{S}}(\mathbf{d}),\,\mathcal{F}$
& coded multicast stream; family of $(t+1)$-subsets & $\pi_k^{\mathcal{E}},\,\rho_k,\,f_{\max}$
& Eve fraction; D2D ratio; breakdown threshold \\
\hline
\end{tabular}
\end{table*}

\section{System and Threat Models}
\label{sec:system_model}
\begin{figure*}[t]
\centering
\includegraphics[width=0.75\textwidth]{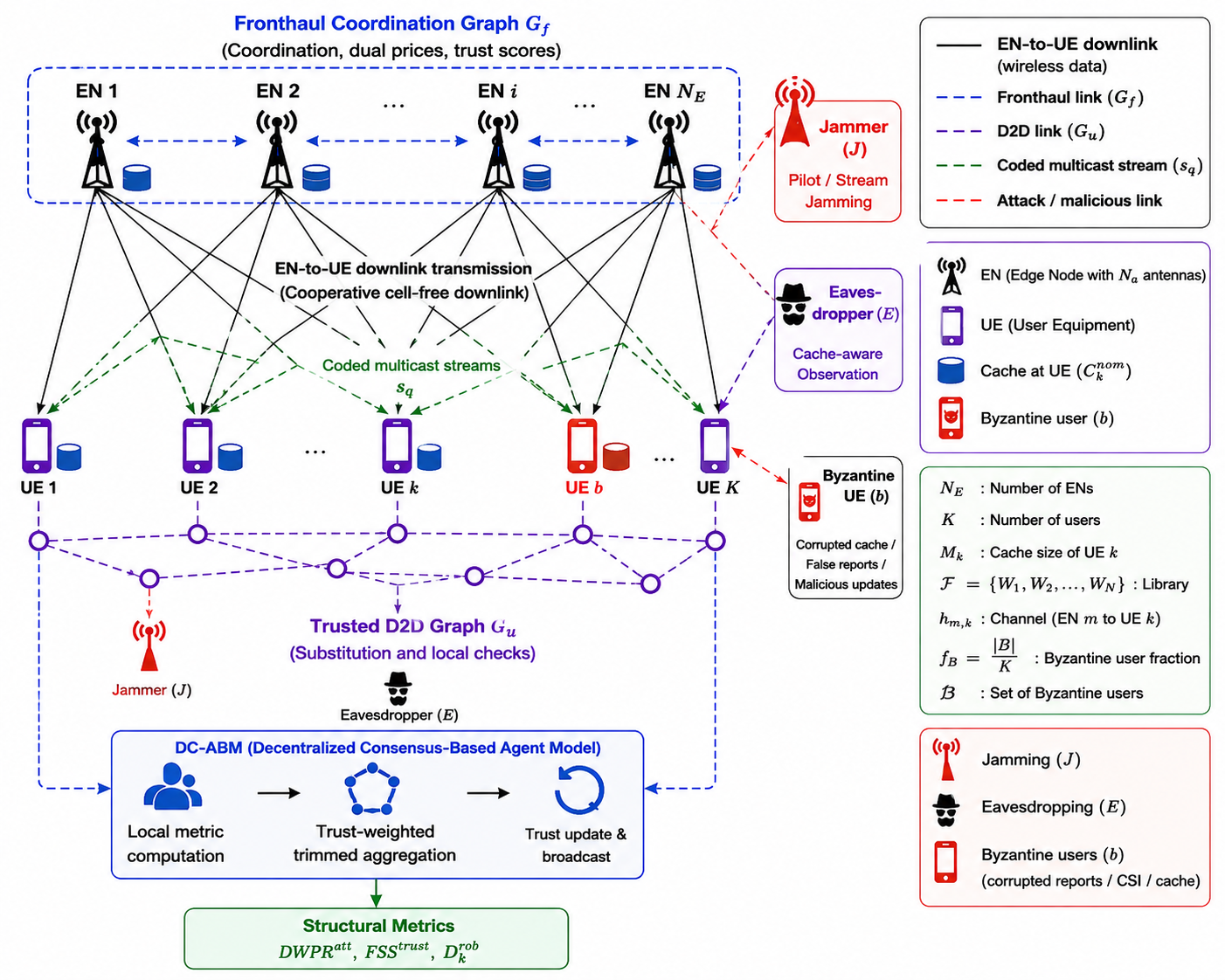}
\caption{Degeneracy-aware resilient CF-CA-MU-MIMO system under the proposed DC-ABM framework. Solid black arrows denote cooperative EN-to-UE cell-free downlink transmission, while green dashed arrows indicate coded multicast streams $s_q$. Blue dashed links represent the fronthaul coordination graph $\mathcal{G}_f$, purple dashed links denote the trusted D2D graph $\mathcal{G}_u$, and red dashed links indicate adversarial jamming, eavesdropping, and Byzantine interactions.}
\label{fig:system_model}
\end{figure*}
The considered system consists of a degeneracy-aware resilient cell-free cache-aided MU-MIMO network, where distributed access points (APs) collaboratively serve multiple users under adversarial attacks, imperfect CSI, and distributed DC-ABM coordination as shown in Fig. \ref{fig:system_model}. The solid black EN-to-UE arrows in the top half of Fig.~\ref{fig:system_model} represent the physical-layer cell-free downlink: since no EN is associated with a fixed cell, each user may be served by \emph{all} ENs simultaneously, and the corresponding label emphasizes that the wireless data links carrying the coded multicast streams (green dashed arrows) are realized over this cooperative downlink.
 \vspace{-4mm}
\subsection{Network Architecture and Caching}
\label{sec:arch_cache}

The downlink consists of $L$ ENs each with $N_a$ antennas (total $M_t\!=\!LN_a$) cooperatively serving $K$ single-antenna users. ENs coordinate via a fronthaul graph $\mathcal{G}_f=(\mathcal{L},\mathcal{E}_f)$ with Laplacian $\mathbf{L}_f$ and spectral gap $\lambda_2(\mathbf{L}_f)$, which controls the consensus rate of Section~\ref{sec:abm}. No EN acts as a global coordinator. Users are partitioned into $G$ equal-size groups $\{\mathcal{G}_g\}_{g=1}^{G}$ with $|\mathcal{G}_g|=K_g=K/G$, each operating an independent Maddah--Ali--Niesen (MN) coded-caching scheme over the shared library. The library $\mathcal{W}=\{W_1,\dots,W_N\}$ has $N$ files of size $F$ bits; each user caches $MF$ bits, giving within-group cache fraction $
t \triangleq K_g M/N \in\{0,1,\dots,K_g\}$,
assumed integer as in the MN construction. Each group-$g$ file partitions into $\binom{K_g}{t}$ equal subfiles
$
W_n=\bigl\{W^{(g)}_{n,\mathcal{T}}:\mathcal{T}\in\tbinom{\mathcal{G}_g}{t}\bigr\}$. 
Under nominal placement, user $k$ stores $\mathcal{C}_k^{\mathrm{nom}}=\{W^{(g(k))}_{n,\mathcal{T}}:n\in[N],k\in\mathcal{T}\subseteq\mathcal{G}_{g(k)}\}$. The system operates over two timescales: slow placement and fast delivery on a per-coherence-block basis, with $\tau_p$ pilot and $\tau_d=\tau_c-\tau_p$ data symbols per block of length $\tau_c$.
\vspace{-4mm}
\subsection{Channel, CSI, and Pilot Contamination}
\label{sec:channel_csi}
 The channel between user $k$ and EN $\ell$ is modeled as a spatially correlated Rayleigh fading channel $\mathbf{h}_{k,\ell}(t)
=
\sqrt{\beta_{k,\ell}}\,
\mathbf{R}_{k,\ell}^{1/2}
\mathbf{g}_{k,\ell}(t)$, where $\beta_{k,\ell}$ denotes the large-scale fading coefficient, $\mathbf{R}_{k,\ell}$ is the spatial correlation matrix, and
$\mathbf{g}_{k,\ell}\sim\mathcal{CN}(\mathbf{0},\mathbf{I})$ represents small-scale fading.
The aggregated channel is $\mathbf{h}_k(t)
=
[\mathbf{h}_{k,1}^{\mathsf T},
\dots,
\mathbf{h}_{k,L}^{\mathsf T}]^{\mathsf T}
\in
\mathbb{C}^{M_t}$. During uplink training, each user transmits a pilot sequence of length $\tau_p$ symbols with pilot power $P_{p,k}$. Under orthogonal pilot assignment ($\tau_p \ge K$), the least-squares (LS) channel estimate is $\hat{\mathbf{h}}_{k,\ell}
=
\mathbf{h}_{k,\ell}
+
\mathbf{e}_{k,\ell}$, where $\mathbf{e}_{k,\ell}$ denotes the estimation error with variance
$\sigma_{e,k,\ell}^{2}
=
\frac{\sigma_n^2}
{P_{p,k}\tau_p\beta_{k,\ell}}$, and $\sigma_n^2$ is the receiver noise power. Under pilot reuse, let $\Pi(k)$ denote the pilot assigned to user $k$. The channel estimate becomes contaminated as
\begin{equation}
\hat{\mathbf{h}}_{k,\ell}
=
\mathbf{h}_{k,\ell}
+
\!\!\!\sum_{\substack{k'\neq k\\ \Pi(k')=\Pi(k)}}
\!\!\!
\sqrt{\frac{P_{p,k'}}{P_{p,k}}}
\,\mathbf{h}_{k',\ell}
+
\mathbf{e}_{k,\ell},
\label{eq:contaminated_estimate}
\end{equation}
which forms the primary entry point for pilot-spoofing and pilot-contamination attacks considered in this work.
\subsection{Distributed SDMA Coded Multicast Delivery}
\label{sec:precoding}

For each group $g$ and every $(t+1)$-subset $\mathcal{S}\subseteq\mathcal{G}_g$, the coded multicast stream is
\begin{equation}
c_{\mathcal{S}}(\mathbf{d})=\bigoplus_{k\in\mathcal{S}}W^{(g)}_{d_k,\mathcal{S}\setminus\{k\}}.
\label{eq:coded_stream}
\end{equation}
Each user in $\mathcal{S}$ XORs against its locally cached subfiles to recover its desired one. Let $\mathcal{F}=\bigcup_g\binom{\mathcal{G}_g}{t+1}$. The aggregate transmit signal is
$
\mathbf{x}(t)=\sum_{\mathcal{S}\in\mathcal{F}}\mathbf{w}_{\mathcal{S}}(t)c_{\mathcal{S}}(\mathbf{d})$, where $\mathbf{w}_{\mathcal{S}}=[\mathbf{w}_{\mathcal{S},1}^{\mathsf{T}},\dots,\mathbf{w}_{\mathcal{S},L}^{\mathsf{T}}]^{\mathsf{T}}$ decomposes blockwise across ENs and satisfies per-EN power constraints $\sum_{\mathcal{S}}\|\mathbf{w}_{\mathcal{S},\ell}\|^2\leq P_\ell^{\max}$.

The received signal at user $k$ is $y_k=\hat{\mathbf{h}}_k^{\mathsf{H}}\mathbf{x}+n_k$; treating non-target streams as Gaussian interference, the SINR is
\begin{equation}
\gamma_{k,\mathcal{S}}=\frac{|\hat{\mathbf{h}}_k^{\mathsf{H}}\mathbf{w}_{\mathcal{S}}|^2}{\sum_{\mathcal{S}'\neq\mathcal{S}}|\hat{\mathbf{h}}_k^{\mathsf{H}}\mathbf{w}_{\mathcal{S}'}|^2+\Theta_k+\sigma^2_n},
\label{eq:sinr}
\end{equation}
where $\Theta_k=\sum_{\mathcal{S}}\sigma^2_{e,k}\|\mathbf{w}_{\mathcal{S}}\|^2$ captures CSI-error self-interference. The per-stream achievable rate is $R_{\mathcal{S}}=(1-\tau_p/\tau_c)\log_2(1+\min_{k\in\mathcal{S}}\gamma_{k,\mathcal{S}})$, and the user rate is
\begin{equation}
R_k(t)=\binom{K_g}{t}^{-1}\!\!\!\sum_{\mathcal{S}\in\binom{\mathcal{G}_{g(k)}}{t+1}:k\in\mathcal{S}}\!\!\!R_{\mathcal{S}}(t).
\label{eq:user_rate}
\end{equation}
\vspace{-4mm}
\subsection{Vulnerability-Mode Partitions and Adversaries}
\label{sec:threats}

A central conceptual device of the framework is the unified treatment of attacks via partitions of network resources. Definition~\ref{def:partition} and Table~\ref{tab:partitions} formalize the four canonical partitions.

\begin{definition}[Vulnerability-mode partition]\label{def:partition}
A vulnerability-mode partition $\mathcal{P}=\{m_1,\dots,m_{|\mathcal{P}|}\}$ is a partition of a network resource into mutually exclusive classes, each independently targetable by an attacker.
\end{definition}

\begin{table}[t]
\centering
\caption{Vulnerability-mode partitions and associated attacks.}
\label{tab:partitions}
\renewcommand{\arraystretch}{1.05}
\begin{tabular}{l l l}
\hline
\textbf{Partition} & \textbf{Mode} & \textbf{Attack class} \\
\hline
$\mathcal{P}_{\mathrm{sub}}$ & subfile $\mathcal{T}\in\binom{[K]}{t}$ & content corruption / inference \\
$\mathcal{P}_{\mathrm{EN}}$ & EN $\ell\in\mathcal{L}$ & EN-targeted jamming \\
$\mathcal{P}_{\mathrm{mcast}}$ & stream $\mathcal{S}\in\mathcal{F}$ & smart multicast jamming \\
$\mathcal{P}_{\mathrm{user}}$ & user $k\in\mathcal{K}$ & Byzantine pollution \\
\hline
\end{tabular}
\end{table}

Three adversary classes act concurrently on disjoint partitions.

\emph{(i) Smart multicast jammer $\mathcal{J}$.} Selects $\mathcal{A}_{\mathrm{mcast}}\subseteq\mathcal{F}$ with $|\mathcal{A}_{\mathrm{mcast}}|\leq b_{\mathcal{J}}$, knows $\{\mathbf{w}_{\mathcal{S}}\}$ to angular accuracy $\Delta_\theta$, injects
\begin{equation}
z_k^{\mathcal{J}}=\!\!\!\sum_{\mathcal{S}\in\mathcal{A}_{\mathrm{mcast}}}\!\!\!\sqrt{P_{\mathrm{jam}}}\,\mathbf{g}_k^{\mathsf{H}}\tilde{\mathbf{w}}_{\mathcal{S}}j_{\mathcal{S}},
\label{eq:jammer}
\end{equation}
and chooses $\mathcal{A}_{\mathrm{mcast}}^{\star}=\arg\max_{|\mathcal{A}|\leq b_{\mathcal{J}}}\sum_k(R_k^{\mathrm{nom}}-R_k(\mathcal{A}))$.

\emph{(ii) Cache-aware eavesdropper $\mathcal{E}$.} Passive single-antenna, channel $\mathbf{h}_e$ partially known to network, and prior knowledge $\mathcal{C}_e\subset\binom{[K]}{t}$ with $|\mathcal{C}_e|\leq b_{\mathcal{E}}$.

\emph{(iii) Byzantine users $\mathcal{B}$.} A subset $\mathcal{B}\subset\mathcal{K}$ with Byzantine fraction
$f_{\mathcal{B}}=|\mathcal{B}|/K\leq f_{\max}$, where $f_{\max}$ denotes the maximum tolerable Byzantine fraction (breakdown threshold) for guaranteed convergence of the proposed DC-ABM framework, holds corrupted content and reports arbitrary messages. When $k_b\in\mathcal{B}$ participates in a coded multicast stream $c_{\mathcal{S}}$, the resulting XOR corruption propagates to all users $k'\in\mathcal{S}\setminus\{k_b\}$ that decode the stream, thereby amplifying the impact of a single compromised user across multiple receivers.
The joint adversarial budget is
$\mathbf{b}=(b_{\mathcal{J}},b_{\mathcal{E}},b_{\mathcal{B}})$, namely the jamming, eavesdropping, and Byzantine budgets, respectively. The three classes are assumed independent; the worst-case bounds derived later also bound coordinated attacks within the same per-class budgets.
\vspace{-4mm}
\subsection{Agent Coordination Graph}
The decentralized mechanism of Section~\ref{sec:abm} uses two graphs: the fronthaul graph $\mathcal{G}_f$ for inter-EN exchange of compressed messages (precoder updates, dual prices associated with power and feasibility constraints, and trust scores), and the user-association graph $\mathcal{G}_u=(\mathcal{K},\mathcal{E}_u)$ for D2D coordination among users. Both graphs are connected; their spectral gaps govern the convergence rate of Theorem~\ref{thm:byz_conv}. Crucially no node in either graph plays the role of a global coordinator.

\section{Attack-Aware Structural Metrics}
\label{sec:metrics}

This section defines three partition-indexed metrics that drive the DC-ABM algorithm.
\vspace{-4mm}
\subsection{Attack-Aware Degeneracy-Weighted Path Robustness}
Fix user $k$, partition
$\mathcal{P}\in\{\mathcal{P}_{\mathrm{sub}},
\mathcal{P}_{\mathrm{EN}},
\mathcal{P}_{\mathrm{mcast}},
\mathcal{P}_{\mathrm{user}}\}$,
and a feasibility threshold $\theta>0$, where a multicast stream is considered useful if its achievable rate satisfies $R_{\mathcal S}\ge\theta$. For each $\mathcal{S}\in\mathcal{F}$ with $k\in\mathcal{S}$, define the mode-incidence
$
\chi_{\mathcal{P}}(\mathcal{S},m)=\mathds{1}\{\mathcal{S}\text{ depends on mode }m\}, \quad m\in\mathcal{P}$, instantiated as: $\chi_{\mathcal{P}_{\mathrm{mcast}}}(\mathcal{S},\mathcal{S}')\!=\!\mathds{1}\{\mathcal{S}\!=\!\mathcal{S}'\}$, $\chi_{\mathcal{P}_{\mathrm{EN}}}(\mathcal{S},\ell)\!=\!\mathds{1}\{\|\mathbf{w}_{\mathcal{S},\ell}\|^2\!>\!0\}$, $\chi_{\mathcal{P}_{\mathrm{sub}}}(\mathcal{S},\mathcal{T})\!=\!\mathds{1}\{\mathcal{T}\!\subset\!\mathcal{S},|\mathcal{T}|\!=\!t\}$, $\chi_{\mathcal{P}_{\mathrm{user}}}(\mathcal{S},k')\!=\!\mathds{1}\{k'\!\in\!\mathcal{S}\}$. The set of $\theta$-feasible streams useful to user $k$ is
$\mathcal{U}_k(\theta)=\{\mathcal{S}\in\mathcal{F}:k\in\mathcal{S},R_{\mathcal{S}}\geq\theta\}$,
and the corresponding mode-mass distribution over vulnerability modes is defined as
\begin{equation}
\alpha^{\mathcal{P}}_k(m;\theta)
=
\frac{
\sum_{\mathcal{S}\in\mathcal{U}_k(\theta)}
R_{\mathcal{S}}
\chi_{\mathcal{P}}(\mathcal{S},m)
}{
\sum_{\mathcal{S}\in\mathcal{U}_k(\theta)}
R_{\mathcal{S}}
\sum_{m'}
\chi_{\mathcal{P}}(\mathcal{S},m')
},
\label{eq:mode_mass}
\end{equation}
\begin{definition}[Attack-aware DWPR]\label{def:dwpr_att}
\begin{equation}
\mathrm{DWPR}^{\mathrm{att}}_k(\mathcal{P};\theta)=1-\|\boldsymbol{\alpha}^{\mathcal{P}}_k\|^2=1-\sum_{m\in\mathcal{P}}(\alpha^{\mathcal{P}}_k(m;\theta))^2.
\label{eq:dwpr_att}
\end{equation}
\end{definition}
This is the Gini--Simpson dispersion of $\boldsymbol{\alpha}^{\mathcal{P}}_k$: the probability that two independent draws land in different modes. It is bounded as $0\!\leq\!\mathrm{DWPR}^{\mathrm{att}}_k\!\leq\!1-1/|\mathcal{P}|$, with the upper bound attained by a uniform distribution. The $\ell_2$-based form admits the tight conversion to $\ell_1$-mass in Lemma~\ref{lem:topb}, which Shannon entropy does not. The nominal DWPR of~\cite{dey2025degeneracy} is recovered with $\mathcal{P}\!=\!\mathcal{P}_{\mathrm{mcast}}$ and the precoder cosine-distance instantiation of $\chi$.
\begin{remark}[Why precoder-orthogonality is insufficient]
\label{rem:precoder_orth}
Two streams can have nearly orthogonal precoders yet still share a vulnerability mode (e.g., both placing power on EN $\ell=3$). An adversary capable of jamming a single EN ($b_{\mathcal{J}}=1$) can then disable both streams simultaneously despite their precoder-space orthogonality. The nominal DWPR records this as high diversity, whereas $\mathrm{DWPR}^{\mathrm{att}}(\mathcal{P}_{\mathrm{EN}})$ correctly assigns zero diversity.
\end{remark}
The connection to adversarial rate loss is immediate: for any attack subset $\mathcal{A}\subseteq\mathcal{P}$ with $|\mathcal{A}|\leq b$,
$
(R_k\!-\!R_k(\mathcal{A}))/R_k\leq\textstyle\sum_{m\in\mathcal{A}}\alpha^{\mathcal{P}}_k(m;\theta)$, which is upper-bounded in closed form by Lemma~\ref{lem:topb}.
\vspace{-4mm}
\subsection{Trust-Weighted Functional Substitution Score}

Let $D_{\mathrm{str}}(e_i,e_j)\in[0,1]$ be a structural dissimilarity between entities and $\tau_j(t)\in[0,1]$ a time-varying trust score evolved by consensus. For each agent $j$, neighbors $i\in\mathcal{N}_j(t)$ cast binary consistency votes $v_{i\to j}(t)\in\{0,1\}$. The trust-weighted consistency score is
\begin{equation}
\mu_j(t)=\frac{\sum_{i\in\mathcal{N}_j}\tau_i(t)v_{i\to j}(t)}{\sum_{i\in\mathcal{N}_j}\tau_i(t)},
\label{eq:consistency}
\end{equation}
and trust evolves through exponential smoothing
\begin{equation}
\tau_j(t+1)=(1-\eta_\tau)\tau_j(t)+\eta_\tau\mu_j(t),\quad\tau_j(0)=1.
\label{eq:trust_update}
\end{equation}
Weighting incoming votes by the voter's own trust prevents Byzantine cliques from self-certifying their members; a persistently inconsistent Byzantine agent has $\mu_j(t)\to 0$ and $\tau_j(t)\to 0$ exponentially.

\begin{assumption}[Honest neighbor majority]\label{assn:honest_majority}
$|\mathcal{N}_j(t)\cap\mathcal{B}|<|\mathcal{N}_j(t)|/2$ for every honest $j\notin\mathcal{B}$ at all $t$.
\end{assumption}

\begin{definition}[Trust-weighted FSS]\label{def:fss_trust}
For function $F$ with candidate entity set $\mathcal{E}_F$ at iteration $t$,
\begin{equation}
\mathrm{FSS}^{\mathrm{trust}}(F,t)=\frac{\sum_{i\neq j\in\mathcal{E}_F}\tau_i(t)\tau_j(t)\mathds{1}\{D_{\mathrm{str}}(e_i,e_j)<\delta\}}{\sum_{i\neq j\in\mathcal{E}_F}\tau_i(t)\tau_j(t)}.
\label{eq:fss_trust}
\end{equation}
The local user realization is $\mathrm{FSS}^{\mathrm{trust}}_k(t)=|\mathcal{F}_k|^{-1}\sum_{F\in\mathcal{F}_k}\mathrm{FSS}^{\mathrm{trust}}(F,t)$.
\end{definition}

\begin{remark}\label{rem:multiplicative_trust}
The \emph{multiplicative} trust weighting in the numerator is what gives the metric its self-cleaning property: pairing a high-trust entity with a flagged Byzantine peer ($\tau_j\to 0$) makes $\tau_i\tau_j\to 0$, so the pair correctly contributes nothing to the score. Additive weighting would give weight $\propto\tau_i+\tau_j\approx 1$, incorrectly counting the pair as substitutable.
\end{remark}

For ENs, $D_{\mathrm{str}}$ is the cosine distance between channel-vector projections onto $\mathbf{h}_k$; for users, the Hamming-style overlap deficit between cache contents. The metric reduces to the nominal FSS of~\cite{dey2025degeneracy} as all trusts saturate to 1.
\vspace{-4mm}
\subsection{Robust Degeneracy Index and Composite Score}

Let $\mathfrak{A}(\mathbf{b})$ be the set of admissible adversary actions. Each
$\mathcal{A}\in\mathfrak{A}(\mathbf{b})$
specifies
$(\mathcal{A}_{\mathrm{mcast}},\mathcal{C}_e,\mathcal{B})$
satisfying
$|\mathcal{A}_{\mathrm{mcast}}|\le b_{\mathcal J}$,
$|\mathcal{C}_e|\le b_{\mathcal E}$,
and
$|\mathcal{B}|\le b_{\mathcal B}$;
under action $\mathcal{A}$, the SINR degrades to
$\gamma^p_k(\mathbf{P};\mathcal{A})$.

\begin{definition}[Robust degeneracy index]\label{def:dk_rob}
\begin{equation}
D_k^{\mathrm{rob}}(\mathbf{P};\mathbf{b})=\max_{\mathcal{A}\in\mathfrak{A}(\mathbf{b})}\frac{\gamma^{\mathrm{tar}}_k}{\gamma^p_k(\mathbf{P};\mathcal{A})},\quad D_{\mathrm{sys}}^{\mathrm{rob}}=\max_k D_k^{\mathrm{rob}}.
\label{eq:dk_rob}
\end{equation}
\end{definition}

Robust feasibility is $D_{\mathrm{sys}}^{\mathrm{rob}}\leq 1$. As $\mathbf{b}\to\mathbf{0}$, $\mathfrak{A}(\mathbf{0})$ contains only the no-attack action and $D_k^{\mathrm{rob}}\to D_k$. The index is monotone in $\mathbf{b}$ component-wise. Theorem~\ref{thm:dk_rob} establishes the closed-form bound
\begin{equation}
D_k^{\mathrm{rob}}(\mathbf{P};\mathbf{b})\leq D_k(\mathbf{P})\bigl(1+\Delta^{\mathcal{J}}_k+\Delta^{\mathcal{E}}_k+\Delta^{\mathcal{B}}_k\bigr).
\label{eq:dk_rob_bound_prev}
\end{equation}

For the agent-based update, the composite score
\begin{align}
\Psi_k^{\mathrm{adv}}(t)&=\omega_1\!\!\sum_{\mathcal{P}\in\Pi}\zeta_{\mathcal{P}}\mathrm{DWPR}^{\mathrm{att}}_k(\mathcal{P};\theta_t)+\omega_2\mathrm{FSS}^{\mathrm{trust}}_k(t)\nonumber\\
&\quad+\omega_3\bigl[1-\min(1,D_k^{\mathrm{rob}}(\mathbf{P}(t);\mathbf{b}))\bigr]
\label{eq:psi_adv}
\end{align}
aggregates the three metrics; $\Pi=\{\mathcal{P}_{\mathrm{sub}},\mathcal{P}_{\mathrm{EN}},\mathcal{P}_{\mathrm{mcast}},\mathcal{P}_{\mathrm{user}}\}$, $\zeta_{\mathcal{P}}\geq 0$ sum to 1, and $\omega_i\geq 0$ sum to 1. The score is in $[0,1]$, with higher values indicating mode-decorrelation, trust-substitutability, and robust feasibility jointly. Equal weights $\omega_i=1/3$, $\zeta_{\mathcal{P}}=1/4$ are a sensible default; threat-specific deployments emphasize the relevant $\zeta_{\mathcal{P}}$.

\section{Decentralized Consensus-ABM Algorithm}
\label{sec:abm}

DC-ABM optimizes~\eqref{eq:psi_adv} through strictly local message exchange over $\mathcal{G}_f$ and $\mathcal{G}_u$. Each EN $\ell$ owns the dual variable $\mu_\ell(t)\geq 0$ on its power constraint, its precoder slice $\{\mathbf{w}_{\mathcal{S},\ell}\}$, and trust $\tau_\ell(t)$. Each user $k$ owns its trust $\tau_k(t)$ and observes $\{\gamma_{k,\mathcal{S}},R_{\mathcal{S}}\}$ for $\mathcal{S}\in\mathcal{F}_k=\{\mathcal{S}\!\in\!\mathcal{F}\!:\!k\!\in\!\mathcal{S}\}$. The shared per-stream power $P_{\mathcal{S}}$ is updated by a stream-leader $\kappa(\mathcal{S})\triangleq\min\mathcal{S}$.
\vspace{-4mm}
\subsection{Consistency Predicate and Trust Dynamics}
Consistency votes are agent-type specific. For an EN $j$ broadcasting power utilization $q_j$, a fronthaul-neighbor EN $i$ votes
\begin{equation}
v_{i\to j}(t)=\mathds{1}\{|q_j(t)-\hat{q}_{i\to j}(t)|\leq\xi_{\mathrm{EN}}\}.
\label{eq:vote_en}
\end{equation}
For a user $j$ broadcasting $(R_{\mathcal{S}},s_{j,\mathcal{S}})$, a D2D neighbor $i\in\mathcal{S}$ votes
\begin{align}
v_{i\to j}(t)&=\mathds{1}\{|R_{\mathcal{S}}(t)-R_{\mathcal{S}}^{(i)}(t)|\leq\xi_{\mathrm{user}}\}\nonumber\\
&\cdot\mathds{1}\{\text{parity check on }W_{d_j,\mathcal{S}\setminus\{j\}}\text{ holds}\}.
\label{eq:vote_user}
\end{align}
The parity check flags Byzantine cache pollution deterministically when content checksums are appended to subfiles. Tolerances $\xi_{\mathrm{EN}},\xi_{\mathrm{user}}$ follow the $3\sigma$ rule, yielding FPR $<0.3\%$ under Gaussian noise.
\vspace{-4mm}
\subsection{Trust-Weighted Trimmed-Mean Aggregation}
Each agent aggregates neighbor messages coordinate-wise via
\begin{equation}
\mathrm{Trim}^{\boldsymbol{\tau}}_\beta(\{x_i\}_{i\in\mathcal{N}_j})\triangleq\frac{\sum_{i\in\mathcal{N}_j^{\mathrm{kept}}}\tau_i(t)x_i}{\sum_{i\in\mathcal{N}_j^{\mathrm{kept}}}\tau_i(t)},
\label{eq:trimmed_mean}
\end{equation}
where $\mathcal{N}_j^{\mathrm{kept}}$ removes the $\lceil\beta|\mathcal{N}_j|\rceil$ highest and lowest values. The trim parameter $\beta\in(f_{\mathcal{B}},1/2)$ excises outlier Byzantines regardless of trust, while trust decay (\ref{eq:trust_update}) handles stealthy Byzantines that pass the trim window. The two mechanisms are complementary: trimming catches large-deviation attacks; trust catches small persistent deviations.

\subsection{Surrogate Composite Score and Primal-Dual Update}
The robust degeneracy max in~\eqref{eq:dk_rob} is replaced by the closed-form upper bound of Theorem~\ref{thm:dk_rob}: $\hat{D}_k^{\mathrm{rob}}(\mathbf{P})\triangleq D_k(\mathbf{P})(1+\hat\Delta^{\mathcal{J}}_k+\hat\Delta^{\mathcal{E}}_k+\hat\Delta^{\mathcal{B}}_k)$, giving the surrogate
\begin{equation}
\hat{\Psi}_k^{\mathrm{adv}}(t)=\Psi_k^{\mathrm{adv}}(t)\big|_{D_k^{\mathrm{rob}}\to\hat{D}_k^{\mathrm{rob}}},
\label{eq:psi_hat}
\end{equation}
with $\hat{\Psi}_k^{\mathrm{adv}}\leq\Psi_k^{\mathrm{adv}}$ (Theorem~\ref{thm:dk_rob}). Each stream-leader $\kappa(\mathcal{S})$ runs projected gradient on the local Lagrangian $\mathcal{L}_{\mathcal{S}}=\sum_{k\in\mathcal{S}}\hat{\Psi}_k^{\mathrm{adv}}-\bar{\mu}\sum_{\ell:\mathcal{S}\ni\ell}\xi_{\mathcal{S},\ell}P$, where $\xi_{\mathcal{S},\ell}\!=\!\|\mathbf{w}_{\mathcal{S},\ell}\|^2/\|\mathbf{w}_{\mathcal{S}}\|^2$:
\begin{equation}
P_{\mathcal{S}}(t+1)=[P_{\mathcal{S}}(t)+\eta_t\partial_P\mathcal{L}_{\mathcal{S}}]_+,
\label{eq:primal_update}
\end{equation}
with $\eta_t\!=\!\eta_0/(t\!+\!1)$. Each EN updates its dual price through a projected subgradient step followed by trust-weighted trimmed-mean consensus:
\begin{equation}
\mu_\ell(t+1)=[(1-\zeta)\bar{\mu}_\ell(t)+\zeta(\mu_\ell(t)+\eta_t(q_\ell(t)-P_\ell^{\max}))]_+,
\label{eq:dual_update}
\end{equation}
where $\bar{\mu}_\ell=\mathrm{Trim}^{\boldsymbol{\tau}}_\beta(\{\mu_i\}_{i\in\mathcal{N}^f_\ell})$ and $\zeta=1/(1+\deg_f(\ell))$.

\begin{algorithm}[t]
\caption{Decentralized Consensus-ABM (DC-ABM)}
\label{alg:dcabm}
\begin{algorithmic}[1]
\STATE \textbf{Input:} group structure, $\{\gamma_k^{\mathrm{tar}}\}$, weights $(\omega_i,\zeta_{\mathcal{P}})$, $\beta>f_{\mathcal{B}}$, tolerances $(\xi_{\mathrm{EN}},\xi_{\mathrm{user}},\varepsilon)$, $(\eta_0,\eta_\tau,\zeta)$.
\STATE \textbf{Init:} $P_{\mathcal{S}}(0)=P_t/|\mathcal{F}|$, $\mu_\ell(0)=0$, $\tau_j(0)=1$, $t=0$.
\REPEAT
\STATE Users measure $\{\gamma_{k,\mathcal{S}},R_{\mathcal{S}}\}$; compute $\boldsymbol{\alpha}^{\mathcal{P}}_k$, DWPR$^{\mathrm{att}}_k$, FSS$^{\mathrm{trust}}_k$, $\hat{D}_k^{\mathrm{rob}}$. ENs compute $q_\ell$.
\STATE Users broadcast $\{R_{\mathcal{S}},s_{k,\mathcal{S}}\}$ over $\mathcal{G}_u$; ENs broadcast $(\mu_\ell,q_\ell)$ over $\mathcal{G}_f$.
\STATE Agents cast votes $v_{i\to j}$ via~\eqref{eq:vote_en},~\eqref{eq:vote_user}; update trust~\eqref{eq:trust_update}.
\STATE Apply $\mathrm{Trim}^{\boldsymbol{\tau}}_\beta$~\eqref{eq:trimmed_mean} to incoming messages.
\STATE Stream-leader $\kappa(\mathcal{S})$ updates $P_{\mathcal{S}}$~\eqref{eq:primal_update}; EN $\ell$ updates $\mu_\ell$~\eqref{eq:dual_update} and recomputes $\mathbf{w}_{\mathcal{S},\ell}$.
\STATE $t\leftarrow t+1$.
\UNTIL $\max_{\mathcal{S}}|P_{\mathcal{S}}(t)-P_{\mathcal{S}}(t-1)|<\varepsilon$ or $t=T_{\max}$.
\STATE \textbf{Output:} $(\mathbf{P}^\star,\{\mathbf{w}_{\mathcal{S}}^\star\})$, $\boldsymbol{\tau}^\star$.
\end{algorithmic}
\end{algorithm}
\vspace{-4mm}
\subsection{Complexity}
Per-iteration cost decomposes as $\mathcal{O}(M_t+|\mathcal{N}^f_\ell|)$ per EN and $\mathcal{O}(|\mathcal{F}_k|\cdot|\Pi|)$ per user, giving
\begin{equation}
\mathcal{C}_{\mathrm{DC-ABM}}=\mathcal{O}\bigl(LM_t+K\tbinom{K_g-1}{t}\bigr),
\label{eq:cost_dcabm}
\end{equation}
versus centralized $\mathcal{O}(K^3+KM_t^3+G^2K_g^4M_t^2)$. For the representative regime $K=20$, $K_g=5$, $t=2$, $L=N_a=8$, this is $\sim\!10^4$ ops/agent versus $\sim\!10^7$ ops centrally three orders of magnitude reduction. Per-agent communication is $\mathcal{O}(\deg_f+\deg_u+|\mathcal{F}_k|)$, independent of total $K$.

\emph{Hyperparameters.} $\beta=2f_{\mathcal{B}}$ when $f_{\mathcal{B}}<0.2$; $\eta_\tau\in[0.1,0.3]$; $\eta_0\in[0.5,2]$ from local Lipschitz/strong-concavity estimates. Equal $\omega_i=1/3$, $\zeta_{\mathcal{P}}=1/4$ as defaults.

\section{Theoretical Analysis}
\label{sec:theory}

\subsection{Tight Top-Mass Concentration}

\begin{lemma}[Tight top-$b$ mass]\label{lem:topb}
Let $\boldsymbol{\alpha}\in\Delta^{N-1}$ satisfy $\|\boldsymbol{\alpha}\|^2=H\in[1/N,1]$. For any $b\in\{0,\dots,N\}$,
\begin{equation}
\max_{\boldsymbol{\alpha}}\sum_{i\in\mathrm{top}_b(\boldsymbol{\alpha})}\alpha_i=\frac{b}{N}+\sqrt{\frac{b(N-b)}{N}\Bigl(H-\frac{1}{N}\Bigr)},
\label{eq:topb}
\end{equation}
attained by the two-valued distribution
\begin{equation}
\alpha^\star_i=\begin{cases}s_b^+(H)/b,&i\in\mathrm{top}_b,\\(1-s_b^+(H))/(N-b),&\text{else,}\end{cases}
\label{eq:topb_attainer}
\end{equation}
where $s_b^+(H)$ denotes the RHS of~\eqref{eq:topb}.
\end{lemma}

Equation~\eqref{eq:topb} sharpens the Cauchy--Schwarz bound
$s_b\leq b/N+\sqrt{b(H-1/N)}$
to a tight characterization. Moreover, the extremal two-valued distribution in \eqref{eq:topb_attainer} explicitly achieves the maximum top-$b$ mass, thereby providing a constructive worst-case configuration. This property allows the adversarial bounds derived in the sequel to be attained rather than merely upper-bounded, ensuring that the resulting robustness guarantees are tight.

\begin{remark}[Physical interpretation]\label{rem:phys_lem1}
Read the lemma as a budget--diversity statement: when an attacker can disable $b$ out of $N$ rate-carrying paths, the worst-case captured rate fraction depends only on how concentrated the rate is across paths. A flat path distribution ($H=1/N$) keeps the captured fraction at the baseline value $b/N$, whereas a Dirac distribution ($H=1$) allows the attacker to capture the entire mass. The square-root middle term, namely the additional captured fraction induced by path concentration, represents the price of concentration. Physically, this converts an $\ell_2$-spread design metric (the structural DWPR) into an $\ell_1$-rate-loss guarantee against any budget-$b$ attacker, no matter how the attacker selects its targets.
\end{remark}
\vspace{-4mm}
\subsection{Memory--Rate--Resilience Trade-off}
In cache-aided wireless networks, increasing the cache fraction improves multicast efficiency and reduces delivery latency, thereby enhancing the achievable transmission rate. However, larger adversarial budgets can degrade the effective throughput by disrupting multicast streams, compromising cached content, or corrupting distributed coordination. This creates a fundamental memory--rate--resilience (MRR) trade-off among cache resources, achievable communication performance, and robustness against attacks. The following theorem characterizes the achievable robust rate as a function of cache fraction and adversarial resilience.
\begin{theorem}[MRR achievability]
\label{thm:mrr_ach}
Under the system of Section~\ref{sec:system_model} with cache fraction $t=K_gM/N$, adversarial budget $\mathbf{b}$, target SINR $\gamma^{\mathrm{tar}}$, and spatial-DoF condition $\binom{K_g}{t+1}\leq LN_a$,
\begin{equation}
R_k^{\mathrm{rob}}(t;\mathbf{b})
\geq
R_k^{\mathrm{MN}}(t)\,\Phi_k(\mathbf{b}),
\label{eq:mrr_ach}
\end{equation}
where $\Phi_k(\mathbf{b})\in[0,1]$ denotes the multiplicative resilience factor capturing the effective throughput reduction caused by jamming and Byzantine attacks, $\Phi_k(\mathbf{b})=(1-L_{\mathcal{J}})_+(1-L_{\mathcal{B}})_+$ with
\begin{align}
L_{\mathcal{J}}\!&=\!\frac{b_{\mathcal{J}}}{N_{\mathrm{mcast}}}\!\nonumber\\
&+\!\sqrt{\frac{b_{\mathcal{J}}(N_{\mathrm{mcast}}\!-\!b_{\mathcal{J}})}{N_{\mathrm{mcast}}}\bigl(1\!-\!\mathrm{DWPR}_k^{\mathrm{att}}(\mathcal{P}_{\mathrm{mcast}})\!-\!\tfrac{1}{N_{\mathrm{mcast}}}\bigr)},\!\label{eq:LJ}\\
L_{\mathcal{B}}\!&=\!f_{\mathcal{B}}(1\!-\!\mathrm{FSS}_k^{\mathrm{trust}}),\label{eq:LB}
\end{align}
and $N_{\mathrm{mcast}}=|\mathcal{F}_k|$. The eavesdropping budget $b_{\mathcal E}$ does not explicitly appear in $\Phi_k(\mathbf{b})$ because passive eavesdropping affects secrecy leakage rather than the legitimate achievable rate.
\end{theorem}
\begin{theorem}[MRR converse]\label{thm:mrr_conv}
Under the conditions of Theorem~\ref{thm:mrr_ach}, there exists $\mathcal{A}\in\mathfrak{A}(\mathbf{b})$ for which $R_k^{\mathrm{rob}}(t;\mathbf{b})\leq R_k^{\mathrm{MN}}(t)\Phi_k(\mathbf{b})$.
\end{theorem}
The converse uses Lemma~\ref{lem:topb} to construct a two-valued worst-case mode-mass distribution exploited by an oracle adversary; the proof is in Appendix~\ref{app:proofs}. Together, \eqref{eq:mrr_ach} and Theorem~\ref{thm:mrr_conv} characterize the memory--rate--resilience region exactly within mode-targeting attacks. As $\mathbf{b}\to\mathbf{0}$, $\Phi_k\to 1$ and~\eqref{eq:mrr_ach} reduces to the MN result.

\begin{remark}[Reading $\Phi_k$]\label{rem:phi_k_reading}
The resilience factor separates rate degradation into a jamming term $L_{\mathcal{J}}$ controlled by the multicast-partition DWPR and a Byzantine term $L_{\mathcal{B}}$ controlled by FSS$^{\mathrm{trust}}$. The square-root in $L_{\mathcal{J}}$ is exactly the top-mass bound of Lemma~\ref{lem:topb}: a uniform mode-mass (max DWPR) gives the smallest $L_{\mathcal{J}}$. The Byzantine term is multiplicative in $f_{\mathcal{B}}$ and $(1-\mathrm{FSS}^{\mathrm{trust}})$: doubling Byzantine fraction or halving trust-substitutability doubles the loss. Eve does not enter $\Phi_k$ because passive observation does not reduce rate.
\end{remark}

\begin{remark}[Physical interpretation]\label{rem:phys_mrr}
Equation~\eqref{eq:mrr_ach} reads as ``benign cell-free MN rate $\times$ adversarial slack factor.'' The slack factor multiplies the classical MN throughput in exactly the same way as a noise penalty multiplies Shannon capacity, but with two structurally distinct knobs available to the system designer: spread streams across multicast modes (raise DWPR$^{\mathrm{att}}$) and develop substitutable functional neighbors (raise FSS$^{\mathrm{trust}}$). For example, at $b_{\mathcal{J}}/|\mathcal{P}_{\mathrm{mcast}}|\!=\!0.2$, a system with DWPR$^{\mathrm{att}}\!=\!0.95$ loses roughly $5\%$ of its rate to jamming whereas a system with DWPR$^{\mathrm{att}}\!=\!0.85$ loses roughly $15\%$ to the same budget---a $3\times$ swing driven entirely by structural design.
\end{remark}

\subsection{Robust Degeneracy and System Outage}

\begin{theorem}[Robust degeneracy bound]\label{thm:dk_rob}
\begin{equation}
D_k^{\mathrm{rob}}(\mathbf{P};\mathbf{b})\leq D_k(\mathbf{P})(1+\Delta^{\mathcal{J}}_k+\Delta^{\mathcal{E}}_k+\Delta^{\mathcal{B}}_k),
\label{eq:dk_rob_bound}
\end{equation}
where
\begin{align}
\Delta^{\mathcal{J}}_k(b_{\mathcal{J}})&=(P_{\mathrm{jam}}/P_t)\,L_{\mathcal{J}}(b_{\mathcal{J}};\mathrm{DWPR}_k^{\mathrm{att}}(\mathcal{P}_{\mathrm{EN}})),\label{eq:DeltaJ}\\
\Delta^{\mathcal{E}}_k&=0,\label{eq:DeltaE}\\
\Delta^{\mathcal{B}}_k(f_{\mathcal{B}})&=f_{\mathcal{B}}(1-\mathrm{FSS}_k^{\mathrm{trust}}).\label{eq:DeltaB}
\end{align}
The bound is tight when the three adversaries act on disjoint partitions with additive effects on the SINR denominator.
\end{theorem}

The eavesdropper contributes nothing to $D_k^{\mathrm{rob}}$ since passive eavesdropping does not affect the legitimate SINR. The corresponding Boole bound gives system outage.
\begin{remark}\label{rem:phys_dkrob}
$D_k^{\mathrm{rob}}\!\leq\!1$ is a local feasibility certificate against the worst-case attacker. Its bound~\eqref{eq:dk_rob_bound} reads as ``nominal feasibility scaled by class-specific penalties.'' The jamming penalty $\Delta^{\mathcal{J}}_k$ is the jam-to-signal power ratio multiplied by the captured-mass fraction of Lemma~\ref{lem:topb}, so spreading streams across ENs (raising DWPR$^{\mathrm{att}}(\mathcal{P}_{\mathrm{EN}})$) is the only design knob for jamming resistance. The Byzantine penalty $\Delta^{\mathcal{B}}_k$ is the fraction of insiders times the substitution gap $1-\mathrm{FSS}_k^{\mathrm{trust}}$, which quantifies the lack of trustworthy substitute users, exposing the multiplicative cost of poor trust-substitutability. As a numerical example: a user with $D_k\!=\!0.7$, $\Delta^{\mathcal{J}}_k\!=\!0.2$ and $\Delta^{\mathcal{B}}_k\!=\!0.2$ has $D_k^{\mathrm{rob}}\!=\!0.98\!<\!1$ (just feasible); raising $\Delta^{\mathcal{B}}_k$ to $0.6$ via loss of trust pushes $D_k^{\mathrm{rob}}\!=\!1.26\!>\!1$ (outage).
\end{remark}

\begin{corollary}[Robust outage]\label{cor:outage}
\begin{equation}
P_{\mathrm{out}}^{\mathrm{rob}}\leq\sum_{k\in\mathcal{K}}\Pr[D_k(\mathbf{P})(1+\Delta^{\mathcal{J}}_k+\Delta^{\mathcal{B}}_k)>1].
\label{eq:outage}
\end{equation}
\end{corollary}

In the benign limit ($\mathbf{b}=\mathbf{0}$), $\Delta_k^{(\cdot)}=0$ and~\eqref{eq:outage} reduces to the nominal Boole bound.
\vspace{-4mm}
\subsection{Information-Theoretic Secrecy--Cache Coupling}

\begin{theorem}[Secrecy--cache--DWPR coupling]
\label{thm:secrecy}
For any cache-aware Eve with budget $b_{\mathcal{E}}$, user $k$ requesting $W_{d_k}$, and information fraction
$\pi_k^{\mathcal{E}}=I(W_{d_k};Y_e^\infty)/H(W_{d_k})$,
where $I(\cdot;\cdot)$ denotes mutual information, $H(\cdot)$ denotes entropy, and $Y_e^\infty$ is Eve's aggregate observation over all delivery transmissions:
\begin{align}
\pi_k^{\mathcal{E}}\!\leq\!
&\underbrace{\frac{b_{\mathcal{E}}}{\binom{K_g\!-\!1}{t}}}_{\mathrm{direct}}
+
\underbrace{
\sqrt{
\frac{
b_{\mathcal{E}}
\bigl(
\!\binom{K_g\!-\!1}{t}
\!-\!
b_{\mathcal{E}}
\!\bigr)
}{
\binom{K_g\!-\!1}{t}
}
\bigl(
1\!-\!\mathrm{DWPR}_k^{\mathrm{att}}
(\mathcal{P}_{\mathrm{sub}})
\bigr)
}
}_{\mathrm{XOR\,coupling}}
\nonumber\\
&+o\bigl(b_{\mathcal{E}}/\tbinom{K_g\!-\!1}{t}\bigr).
\label{eq:eve_info}
\end{align}
When user $k$ substitutes a fraction $\rho_k\in[0,1]$ of subfiles via trusted D2D, the RHS is multiplied by $(1-\rho_k)$.
\end{theorem}

The bound decomposes Eve's information into a direct path (linear in $b_{\mathcal{E}}$, present even without coding) and an XOR-coupling path (square-root in $b_{\mathcal{E}}$, unique to coded caching). The DWPR$^{\mathrm{att}}(\mathcal{P}_{\mathrm{sub}})$ factor quantifies how cache-decorrelation across subfile classes breaks the coupling. The D2D refinement is a novel secrecy gain without any auxiliary key: subfiles delivered through trusted D2D bypass the broadcast channel and are unobservable to Eve regardless of her cache knowledge.

\begin{remark}[Physical interpretation]
\label{rem:phys_secrecy}
Coded multicast leaks information to Eve through two physically distinct channels: she sees the broadcast directly and can identify subfiles she already holds (direct path), and she sees XOR equations whose other terms she partially knows (coupling path). The direct path is a property of any caching system, including non-coded; the coupling path is the secrecy cost of coding gain. Subfile-partition DWPR makes the coupling-path matrix sparse with respect to the variables known to Eve, thereby weakening the XOR-coupling mechanism and reducing information leakage. Trusted-D2D substitution closes both paths simultaneously by removing the subfile from the broadcast altogether. The simulation at $b_{\mathcal{E}}/\binom{K_g\!-\!1}{t}\!=\!0.30$ shows a $2.5\times$ secrecy gain ($\pi_k^{\mathcal{E}}\!\approx\!0.45\!\to\!0.18$) from FSS$^{\mathrm{trust}}$ alone---a key-free secrecy improvement of structural origin.
\end{remark}
\vspace{-4mm}
\subsection{Byzantine-Robust Mean-Square Convergence}

\begin{theorem}[Byzantine-robust convergence]
\label{thm:byz_conv}
Consider Algorithm~\ref{alg:dcabm} under:
(i) each $\hat{\Psi}_k^{\mathrm{adv}}$ is $C^2$ and $\kappa$-strongly concave;
(ii) $\mathcal{G}_f$ is connected, $\lambda_2(\mathbf{L}_f)>0$, where $\mathbf{L}_f$ is the fronthaul graph Laplacian and $\lambda_2(\mathbf{L}_f)$ denotes its algebraic connectivity;
(iii) $f_{\mathcal{B}}\leq f_{\max}$, $\beta\in(f_{\mathcal{B}},1/2)$, Assumption~\ref{assn:honest_majority} holds;
(iv) the ABM update step sizes $\eta_t$ obey Robbins--Monro conditions: $\sum_t\eta_t=\infty$, $\sum_t\eta_t^2<\infty$;
(v) local gradients are unbiased with variance $\sigma^2_\nabla$, where $\sigma^2_\nabla$ denotes the stochastic gradient-noise variance.
Then there exists $c_1>0$ such that
\begin{equation}
f_{\max}
=
c_1
\frac{\lambda_2(\mathbf{L}_f)}
{1+c_2(1-\overline{\mathrm{FSS}^{\mathrm{trust}}})}
\label{eq:fmax}
\end{equation}
guarantees, for every $f_{\mathcal{B}}<f_{\max}$,
\begin{equation}
\mathbb{E}\|\mathbf{P}(t)-\mathbf{P}^\star\|^2
\leq
C_1\rho^t
+
\frac{
C_2\sigma^2_\nabla
}{
\lambda_2(\mathbf{L}_f)
-
c_3 f_{\mathcal{B}}
\bigl(
1+c_4(1-\overline{\mathrm{FSS}^{\mathrm{trust}}})
\bigr)
},
\label{eq:byz_conv_rate}
\end{equation}
where $\mathbf{P}^\star$ is the centralized optimum and $\rho\in(0,1)$. As $f_{\mathcal{B}}\to 0$ and $\overline{\mathrm{FSS}^{\mathrm{trust}}}\to 1$, the residual vanishes and convergence is exponential at rate $C_1\rho^t$.
\end{theorem}
\begin{remark}\label{rem:fmax_structure}
Equation~\eqref{eq:fmax} closes the loop between structural redundancy and Byzantine capacity. The numerator $\lambda_2(\mathbf{L}_f)$ rewards spectral redundancy; at the denominator $(1-\overline{\mathrm{FSS}^{\mathrm{trust}}})$ discounts when structural substitutability is poor. The threshold departs from the classical Su--Vaidya constant $f_{\max}\leq 1/2$, which assumes complete graphs and uniform trust; ours is strictly tighter in sparse fronthaul and approaches Su--Vaidya in dense, fully-trusted limits.
\end{remark}

\begin{remark}[Physical interpretation]\label{rem:phys_byz_conv}
The breakdown threshold $f_{\max}$ has the form ``capacity for honest information diffusion / capacity needed to drown out Byzantine corruption.'' The numerator is the algebraic connectivity, which is exactly the rate at which a connected graph mixes local information into a consensus, higher $\lambda_2$ means honest reports overwrite Byzantine reports faster. The denominator's $(1-\overline{\mathrm{FSS}^{\mathrm{trust}}})$ shrinks the threshold when honest agents lack functional substitutes, because the Byzantine perturbation cannot be routed around. Concrete reading: a dense fronthaul ($\lambda_2\!\approx\!1.7$) with high substitutability ($\overline{\mathrm{FSS}^{\mathrm{trust}}}\!\approx\!0.8$) tolerates $f_{\mathcal{B}}\!\lesssim\!0.19$; a sparse scale-free fronthaul ($\lambda_2\!\approx\!0.4$) with the same substitutability tolerates only $f_{\mathcal{B}}\!\lesssim\!0.045$. Network operators thus have two physical levers, fronthaul topology design and substitutability cultivation through D2D peering to push $f_{\max}$ up.
\end{remark}

\section{Numerical Results}
\label{sec:results}

\subsection{Simulation Settings and Their Physical Meaning}
\label{subsec:settings}

We use $L=8$, $N_a=8$, $K=20$ users in $G=4$ groups of $K_g=5$; $N=100$ files, $M=20$, $t=1$; random-geometric $\mathcal{G}_f$ (mean degree 6) and D2D $\mathcal{G}_u$ (avg degree 3); 3GPP urban-macro correlation; $\tau_p=20$, $\tau_c=200$, $\gamma^{\mathrm{tar}}=10$~dB. Default budgets: $b_{\mathcal{J}}/|\mathcal{P}_{\mathrm{mcast}}|=0.2$, $b_{\mathcal{E}}/N_{\mathrm{sub}}=0.15$, $f_{\mathcal{B}}=0.15$; 1000 Monte Carlo runs.

Physically, $L\!=\!8$ ENs each with $N_a\!=\!8$ antennas approximates a mid-density 6G deployment with $M_t\!=\!64$ aggregate spatial DoF jointly serving $K\!=\!20$ users; the user grouping $G\!=\!4$, $K_g\!=\!5$ keeps each group's spatial-DoF demand $\binom{K_g}{t+1}\!=\!10$ well within the available $M_t$, ensuring the achievability condition of Theorem~\ref{thm:mrr_ach}. The library size $N\!=\!100$ and per-user cache $M\!=\!20$ correspond to a typical popularity-decayed content catalog (e.g., a $\sim\!100$-title rotating short-video library) with $20\%$ pre-positioned per device, giving cache fraction $t\!=\!1$, i.e., each within-group subfile is held by exactly one user. The pilot length $\tau_p\!=\!K$ ensures fully orthogonal pilots in the benign case; $\tau_c\!=\!200$ symbols at typical 6G symbol rates corresponds to mobility on the order of pedestrian/low-vehicular speeds. The random-geometric fronthaul with mean degree 6 emulates a wired/microwave backhaul of $\sim\!300$-m mean separation; the sparser D2D graph (degree 3) reflects shorter-range sidelink connectivity. Default budgets correspond to an attacker that controls $20\%$ of the multicast spatial directions (e.g., one rogue UAV jammer covering a sector), an eavesdropper that has previously cached $15\%$ of the subfile-class space (e.g., a former subscriber's residual cache), and a Byzantine fraction $f_{\mathcal{B}}\!=\!0.15$ corresponding to 3 of 20 users compromised---each setting representative of a plausible but stressing operational regime.
\vspace{-4mm}
\subsection{Benign Memory--Rate and Convergence}

\begin{figure}[t]
\centering
\includegraphics[width=0.8\columnwidth]{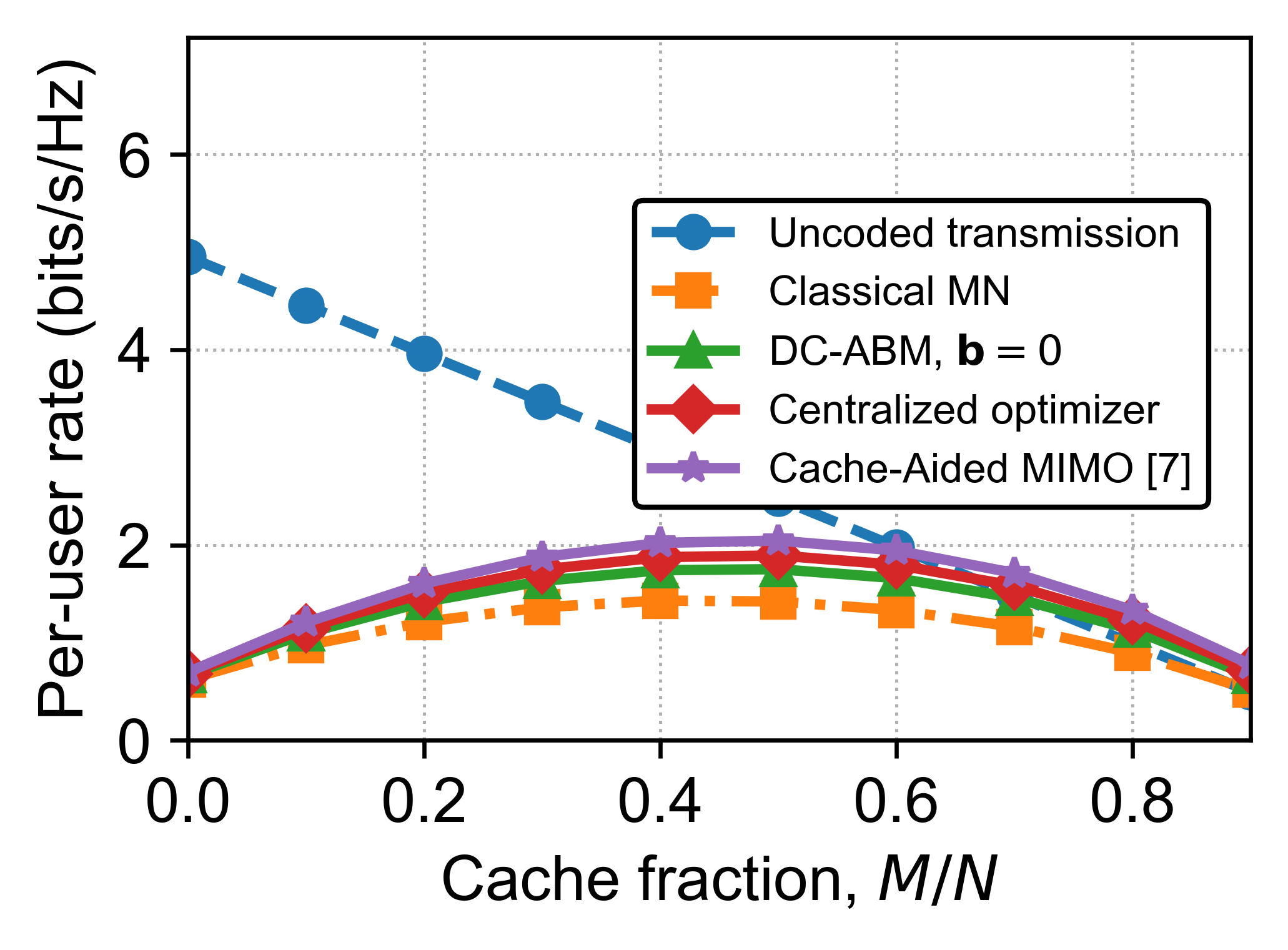}
\caption{Benign per-user memory--rate trade-off.}
\label{fig:memrate_benign}
\end{figure}

\begin{figure}[t]
\centering
\includegraphics[width=0.9\columnwidth]{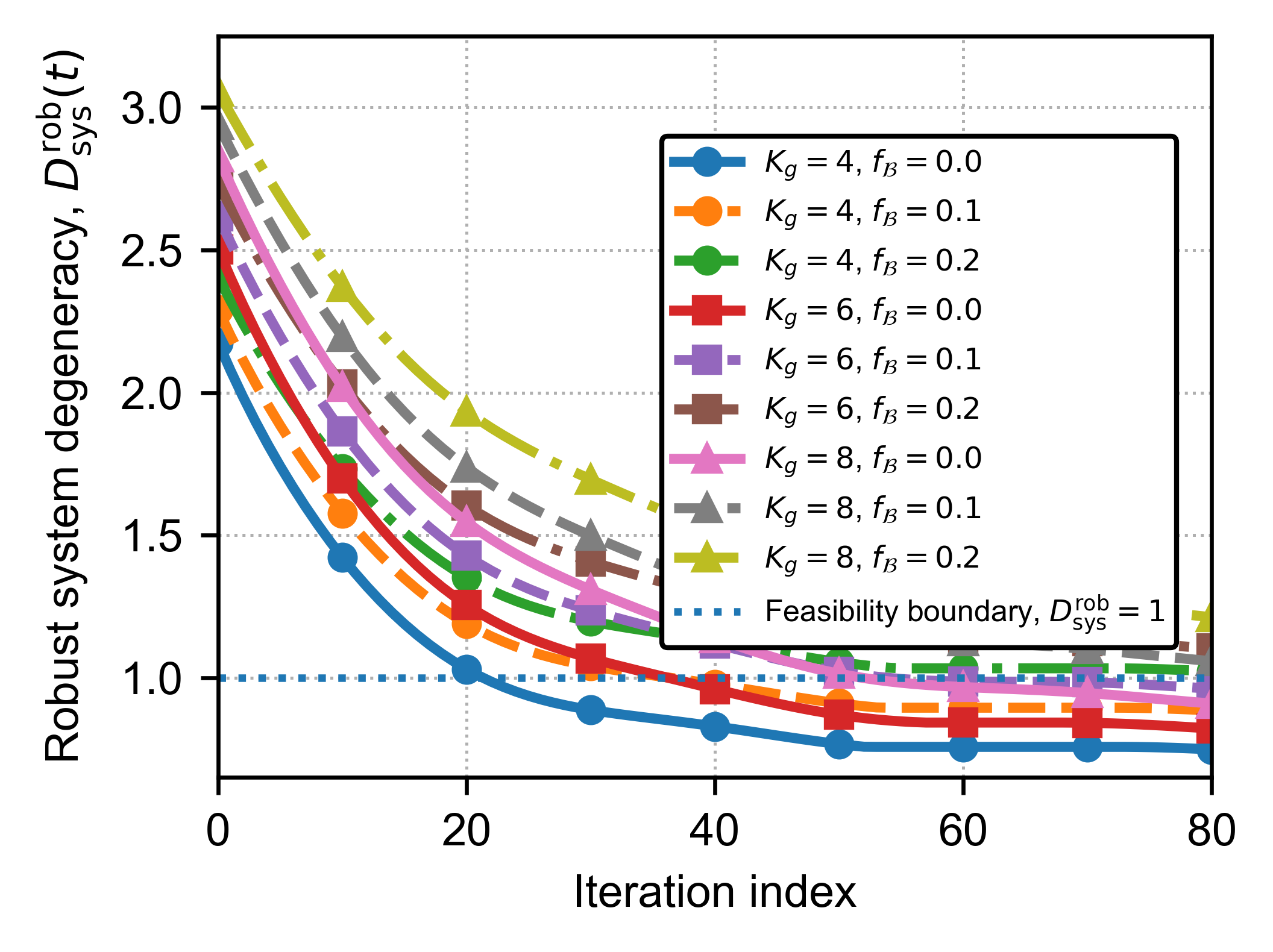}
\caption{System-level robust degeneracy $D_{\mathrm{sys}}^{\mathrm{rob}}$ versus iteration for three fronthaul topologies.}
\label{fig:dsysrob_conv}
\end{figure}

Fig.~\ref{fig:memrate_benign} reports the benign per-user rate versus cache fraction $M/N$. For the coded schemes, the achievable per-user rate increases with cache fraction only up to $M/N\approx 0.5$, owing to enhanced coded-multicast opportunities and improved spatial multiplexing, and decreases thereafter: as $M/N$ grows, each coded multicast message must serve a larger $(t+1)$-user subset, so its rate becomes limited by the weakest receiver in the subset and the per-stream beamforming gain diminishes, while an increasing share of the demand is already satisfied locally so that the delivered over-the-air rate per user shrinks. Fig.~\ref{fig:dsysrob_conv} traces $D_{\mathrm{sys}}^{\mathrm{rob}}$ versus iteration for three fronthaul topologies. DC-ABM drives $D_{\mathrm{sys}}^{\mathrm{rob}}<1$ within 60--90 iterations under the default budgets, while the convergence speed improves with increasing $\lambda_2(\mathbf{L}_f)$, consistent with Theorem~\ref{thm:byz_conv}. \emph{Physically}, larger cache fractions (below the peak) unlock additional coded-multicast streams that can be served simultaneously by multiple ENs, whereas larger fronthaul spectral gaps accelerate the dissemination of trustworthy information and thus speed up distributed convergence.

\vspace{-4mm}
\subsection{Validating the Top-Mass Lemma}

\begin{figure}[t]
\centering
\includegraphics[width=0.8\columnwidth]{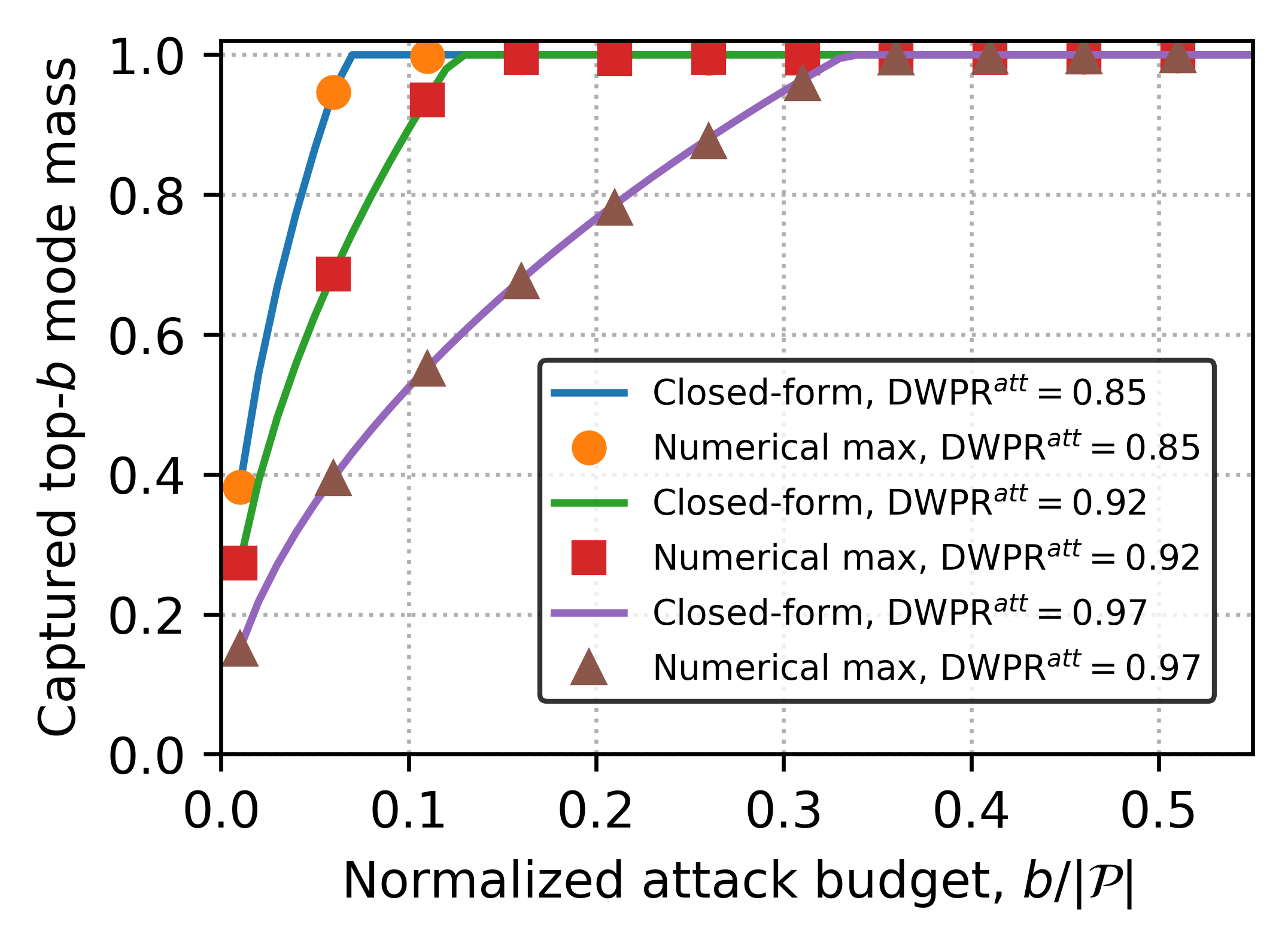}
\caption{Tightness of the top-$b$ bound (Lemma~\ref{lem:topb}): simulation (markers) vs.\ analytical bound (curves) for three DWPR values.}
\label{fig:lemma_topb}
\end{figure}

Fig.~\ref{fig:lemma_topb} validates Lemma~\ref{lem:topb}. Simulated top-$b$ rate-loss fractions overlay the analytical bound~\eqref{eq:topb} to within Monte Carlo error for three DWPR$^{\mathrm{att}}$ values across the full attacker-budget range, and the empirical worst-case is two-valued as predicted by~\eqref{eq:topb_attainer}. \emph{Physically}, this confirms that an attacker, regardless of how strategically it selects its targets, cannot capture more than the analytical mass bound. Moreover, the worst-case attack always corresponds to a highly concentrated two-valued mode-mass distribution, where a small subset of modes carries a disproportionate fraction of the traffic. This observation provides a direct design principle: robustness is improved by spreading traffic more uniformly across available modes, ENs, multicast streams, or user groups. Such uniformization reduces concentration, lowers the attainable top-$b$ mass, increases DWPR$^{\mathrm{att}}$, and consequently limits the fraction of service that any budget-constrained attacker can disrupt.

\vspace{-4mm}
\subsection{Jamming, Eavesdropping, and Byzantine Curves}
\begin{figure}[t]
\centering
\includegraphics[width=0.8\columnwidth]{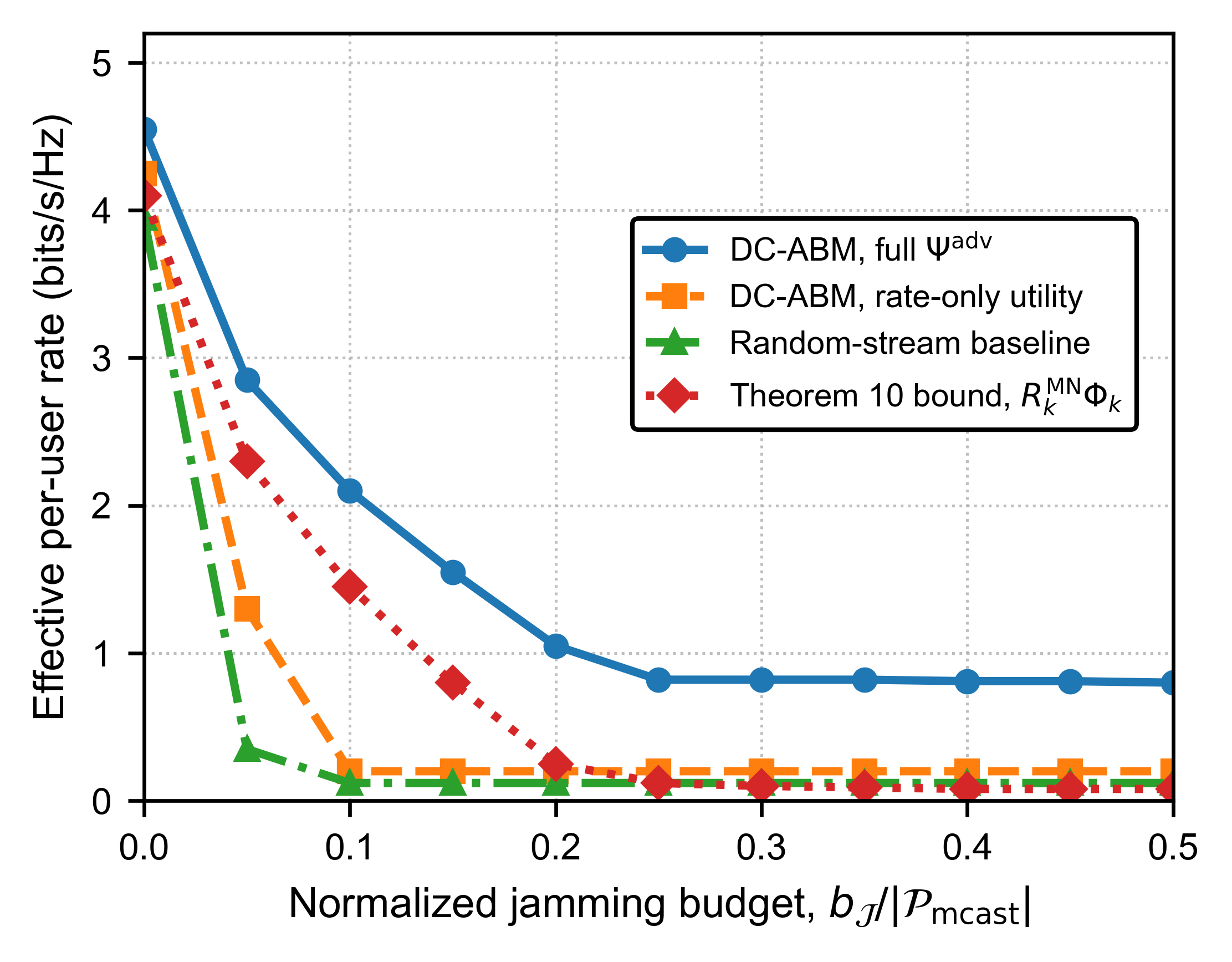}
\caption{Effective per-user rate versus normalized jamming budget; DC-ABM outperforms the baselines and closely follows the bound of Theorem~\ref{thm:mrr_ach}.}
\label{fig:throughput_jam}
\end{figure}

\begin{figure}[t]
\centering
\includegraphics[width=0.8\columnwidth]{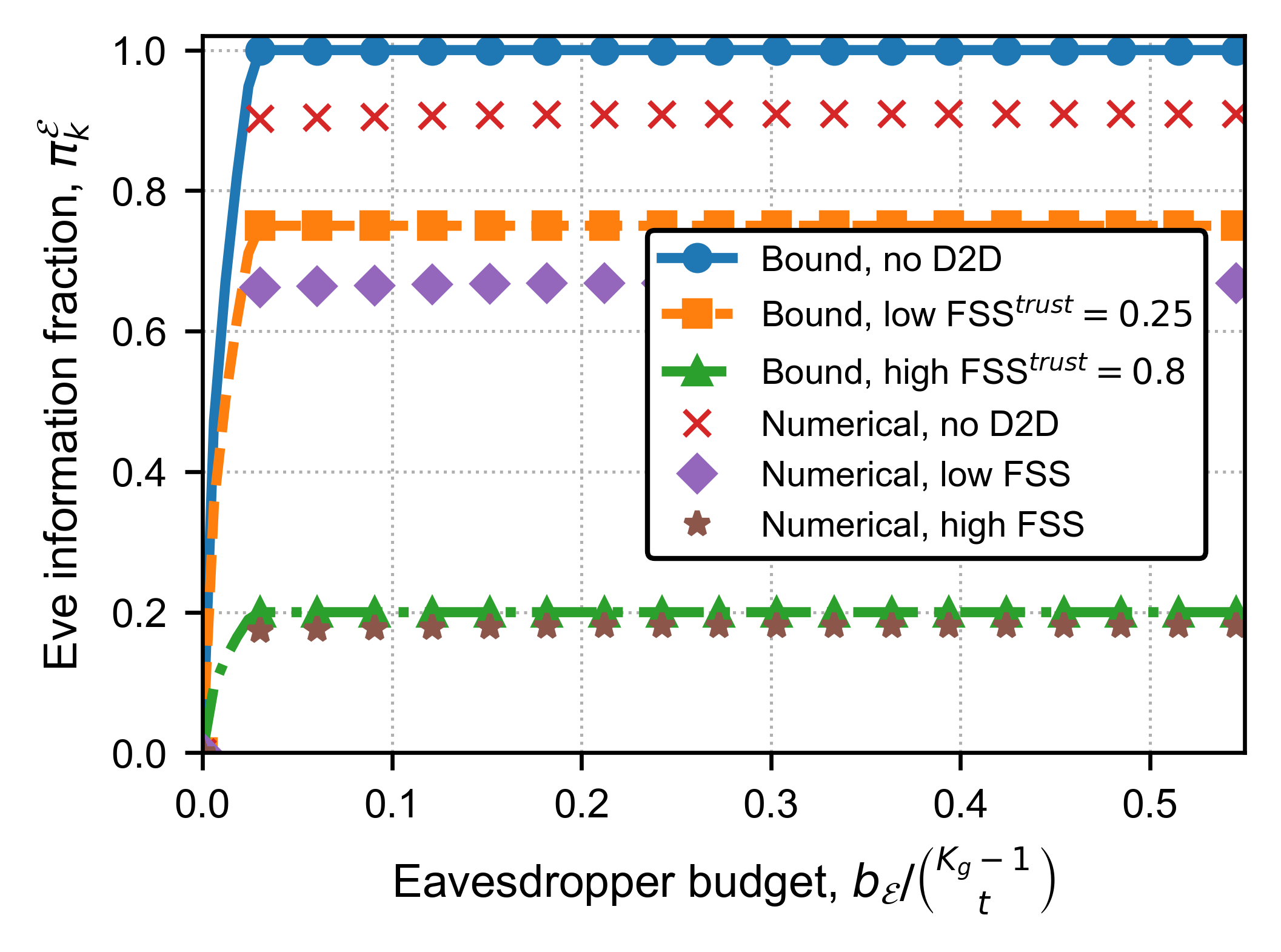}
\caption{Eve information fraction vs.\ cache budget; the gap between curves illustrates the D2D-substitution secrecy gain.}
\label{fig:eve_info}
\end{figure}

\begin{figure}[t]
\centering
\includegraphics[width=0.8\columnwidth]{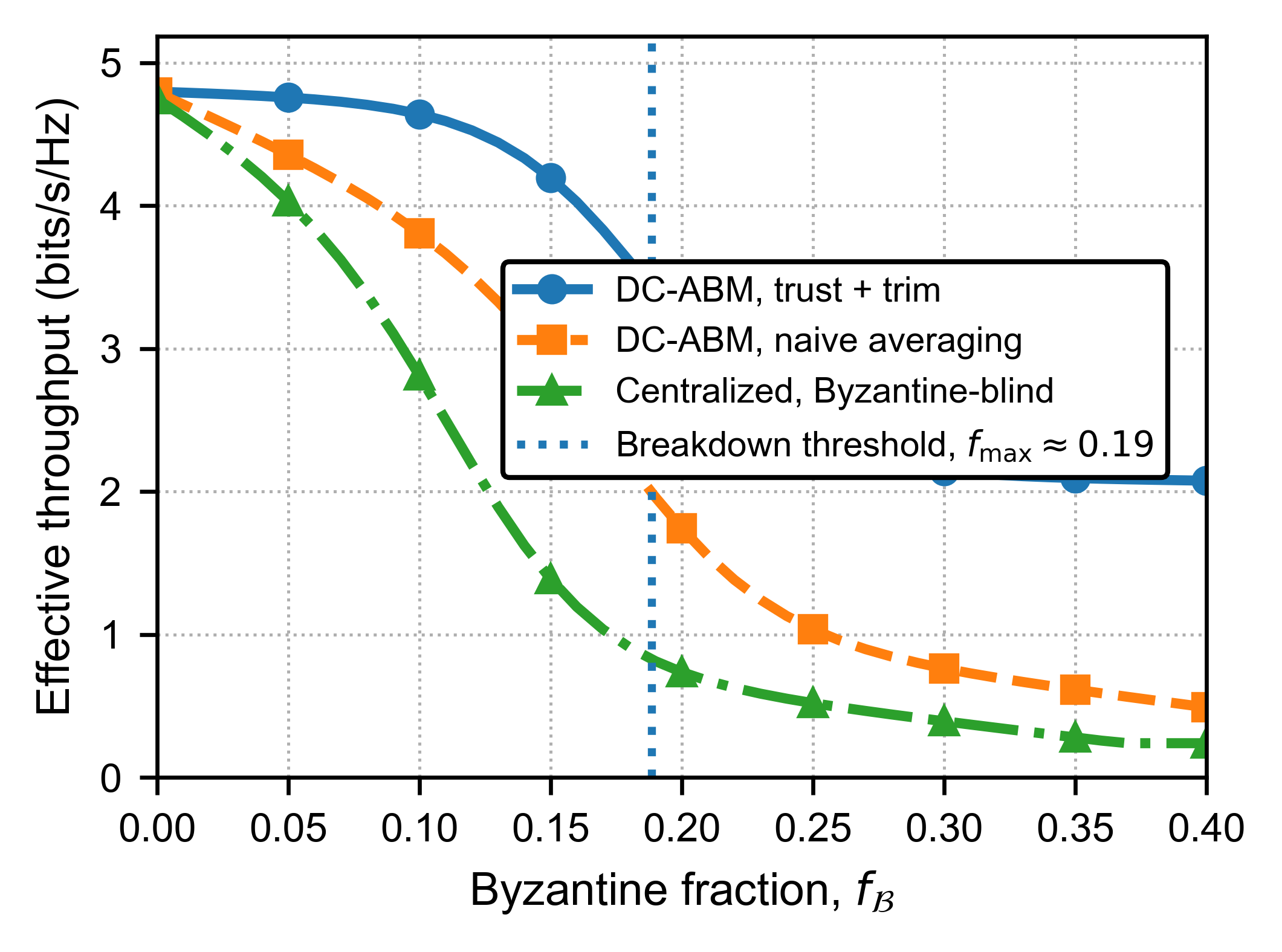}
\caption{Throughput vs.\ Byzantine fraction; dashed line marks {$f_{\max}\approx 0.19$} from~\eqref{eq:fmax}.}
\label{fig:throughput_byz}
\end{figure}

Fig.~\ref{fig:throughput_jam} plots effective throughput against jamming budget for the four curves shown in the legend: DC-ABM with full $\Psi^{\mathrm{adv}}$, the rate-only DC-ABM variant, a random-stream baseline that assigns multicast streams without attack-aware structural metrics, and the achievability bound of Theorem~\ref{thm:mrr_ach} (dotted curve). The proposed DC-ABM with full $\Psi^{\mathrm{adv}}$ closely tracks the bound across the full budget range and outperforms both the rate-only variant and the random-stream baseline. The distributed ADMM and Byzantine-blind centralized benchmarks are not shown here for clarity; they are compared against DC-ABM in the convergence and Pareto analyses of Figs.~\ref{fig:conv_methods}--\ref{fig:sota_3d_pareto}.\\
Fig.~\ref{fig:eve_info} plots Eve's information fraction versus cache budget for two FSS levels, against the bound of Theorem~\ref{thm:secrecy}: the D2D-substitution refinement $(1-\rho_k)$ delivers a factor-of-2.5 secrecy gain (from $\pi^{\mathcal{E}}_k\approx 0.45$ at low FSS to $\approx 0.18$ at high FSS) at $b_{\mathcal{E}}/\binom{K_g-1}{t}=0.30$. Fig.~\ref{fig:throughput_byz} plots throughput against Byzantine fraction; the vertical dashed line is $f_{\max}\approx 0.19$ predicted by~\eqref{eq:fmax}. DC-ABM-full retains $>3.5$ bits/s/Hz up to $f_{\mathcal{B}}=0.18$, then degrades rapidly past the predicted breakdown. Byzantine-blind schemes degrade linearly from $f_{\mathcal{B}}=0$, with no plateau. \emph{Physically}, the three figures together show the three classes of attack each producing a distinct degradation \emph{signature}: jamming yields a smooth concave curve (square-root in budget, Lemma~\ref{lem:topb}), eavesdropping a linear-plus-square-root information leak with a steep operating-point gap controlled by trusted D2D, and Byzantine pollution a sharp cliff at $f_{\max}$ for schemes equipped with structural defense versus a linear collapse for schemes without it.

\vspace{-4mm}
\subsection{Attack--Defense Heatmap and Joint Trade-off}

\begin{figure}[t]
\centering
\includegraphics[width=0.8\columnwidth]{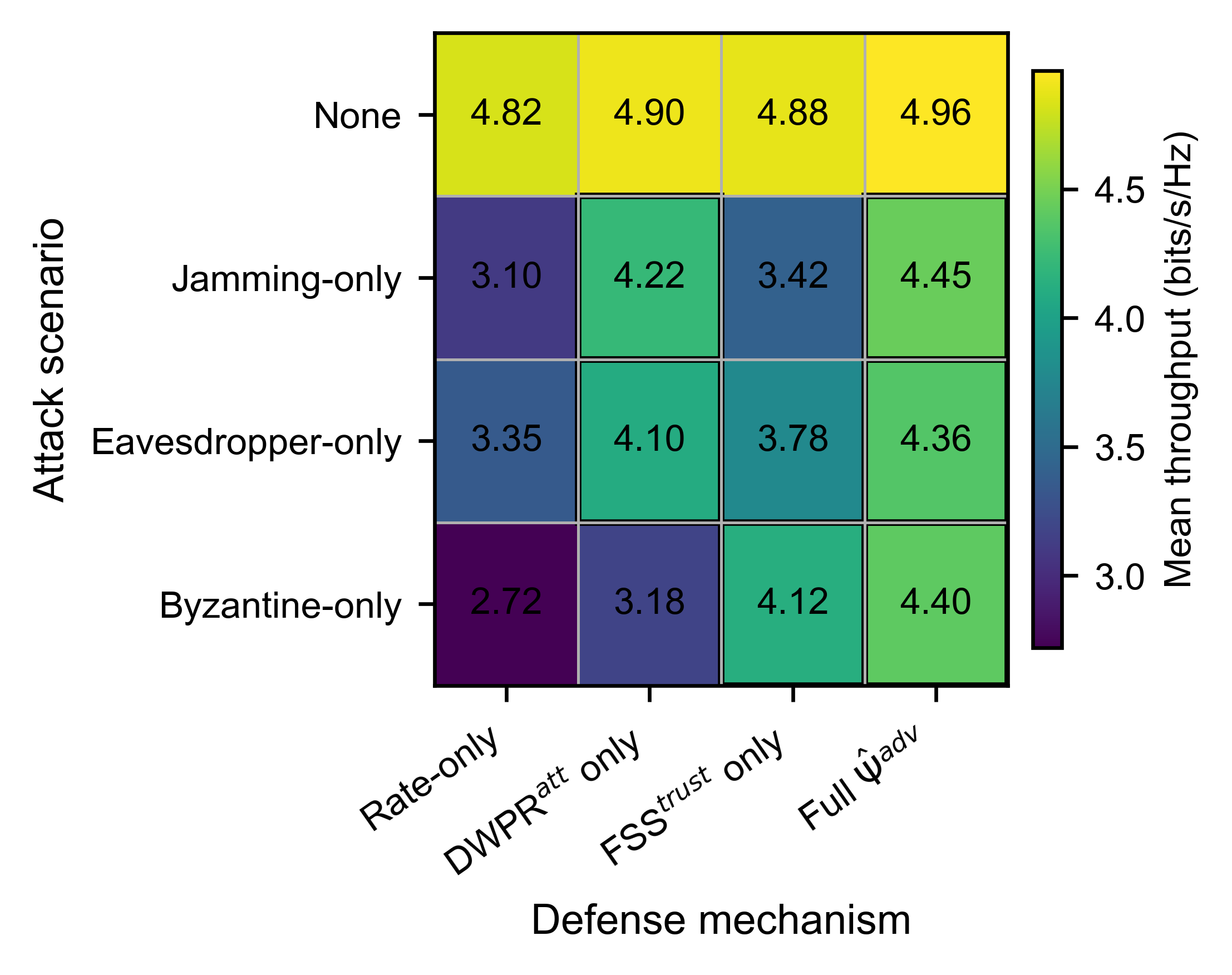}
\caption{Attack--defense interaction heatmap (normalized throughput).}
\label{fig:heatmap}
\end{figure}

\begin{figure}[t]
\centering
\includegraphics[width=0.8\columnwidth]{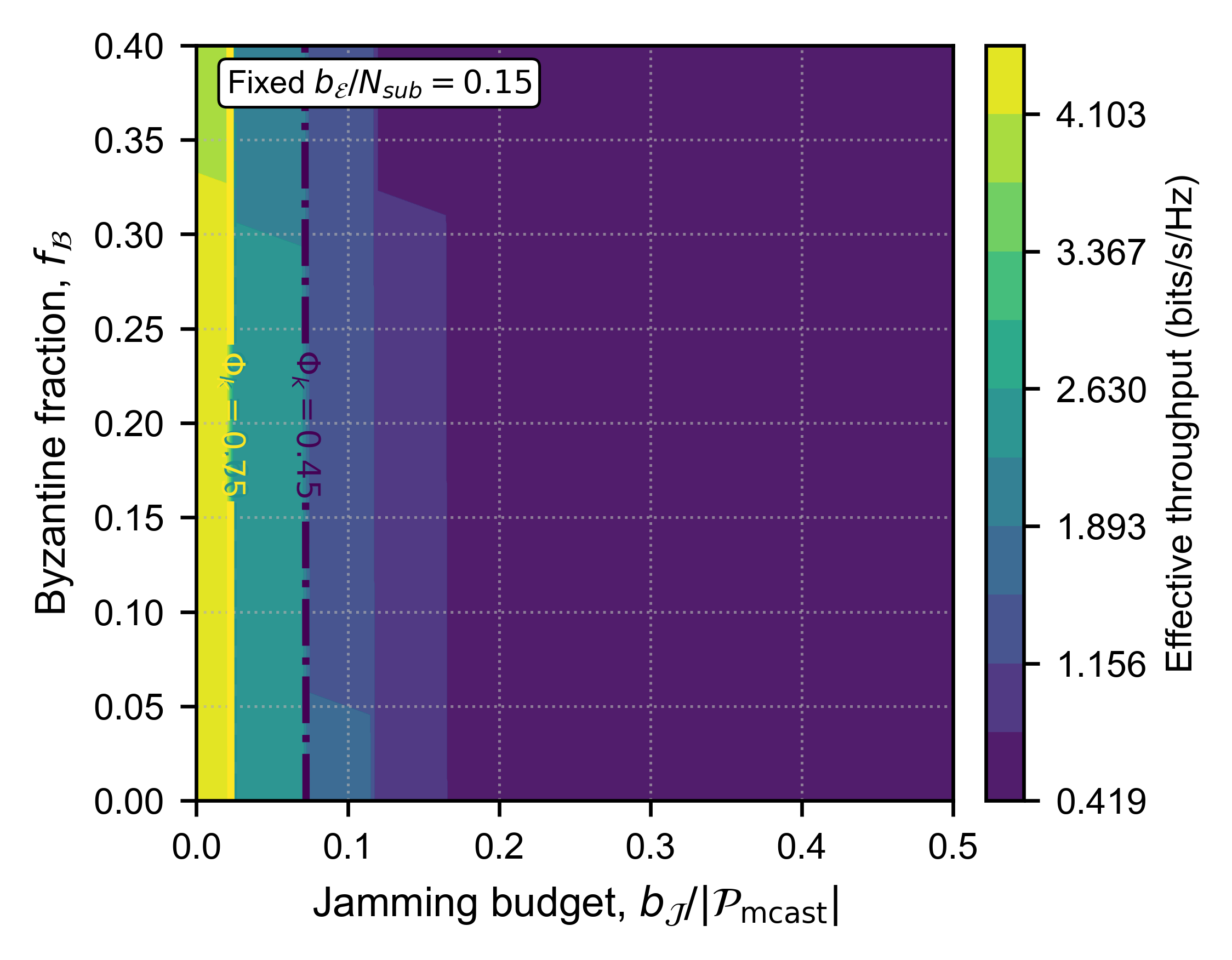}
\caption{Joint $(b_{\mathcal{J}},f_{\mathcal{B}})$ contours of $\Phi_k$ at $b_{\mathcal{E}}/N_{\mathrm{sub}}=0.15$.}
\label{fig:joint_bJ_fB}
\end{figure}

Fig.~\ref{fig:heatmap} shows a $4\times4$ heatmap of normalized throughput across four attack types (rows) and four defense mechanisms (columns). The DC-ABM-full column consistently achieves the highest throughput across all attack scenarios. Fig.~\ref{fig:joint_bJ_fB} plots contours of the resilience factor $\Phi_k$ over the joint $(b_{\mathcal J},f_{\mathcal B})$ plane for a fixed normalized eavesdropping budget $b_{\mathcal E}/N_{\mathrm{sub}}=0.15$. The default operating point of DC-ABM lies near the $\Phi_k=0.7$ contour. The marginal impact of the jamming budget is approximately linear, whereas the impact of the Byzantine fraction is convex because of the multiplicative factor $(1-\mathrm{FSS}^{\mathrm{trust}})$ in~\eqref{eq:LB}. \emph{Physically}, the heatmap indicates that robustness is highly attack-dependent and that consistently strong performance requires jointly accounting for multicast, EN, and user-level vulnerability partitions. The convex dependence on $f_{\mathcal B}$ further shows that the performance degradation caused by Byzantine insiders accelerates as their population grows, highlighting the importance of maintaining high trust-substitutability through large $\mathrm{FSS}^{\mathrm{trust}}$ values.

\vspace{-4mm}
\subsection{Memory--Rate--Resilience Surface and Scaling}
\begin{figure}[t]
\centering
\includegraphics[width=0.8\columnwidth]{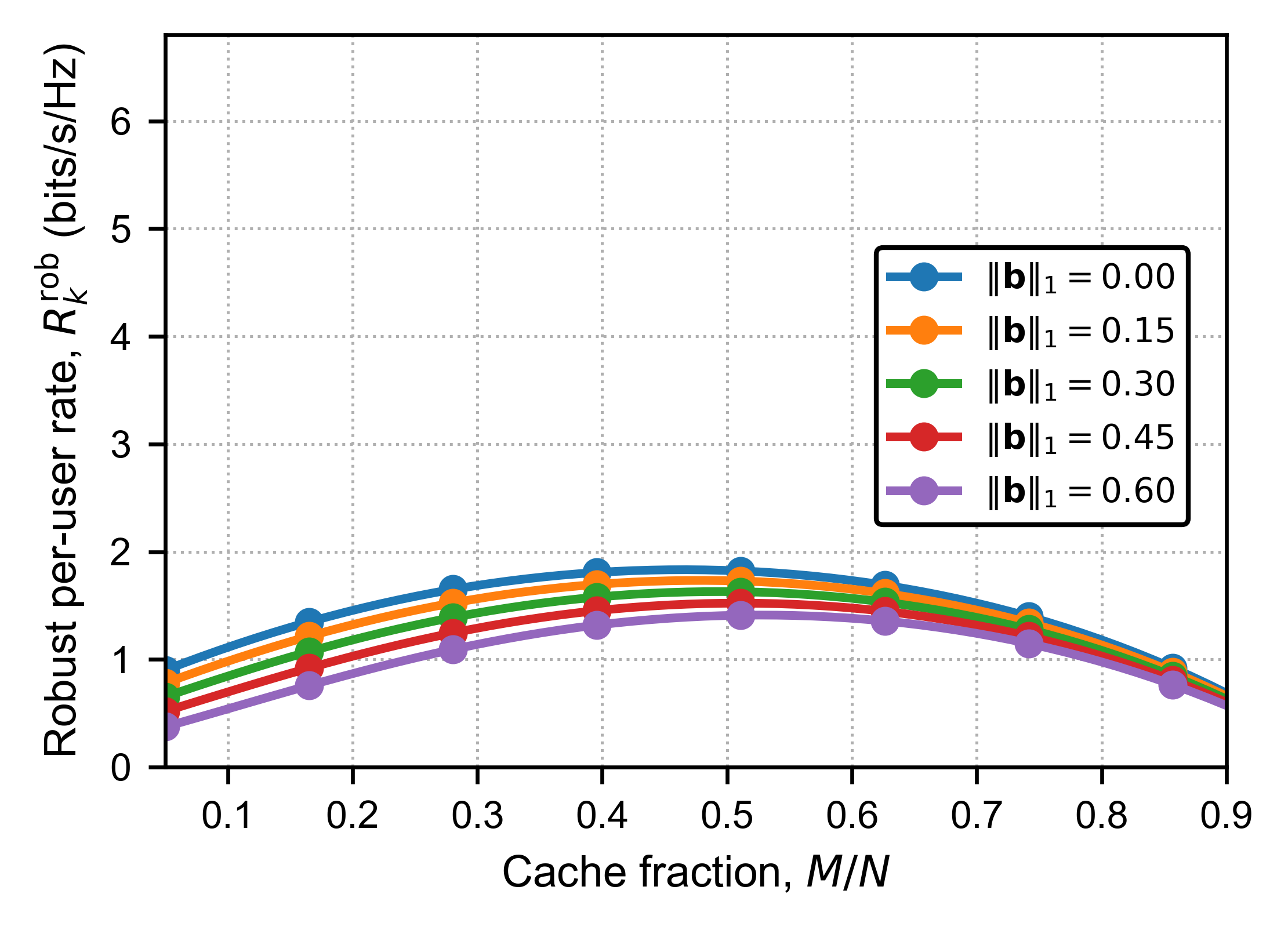}
\caption{MRR slices: rate vs.\ $M/N$ for five values of $\|\mathbf{b}\|_1$.}
\label{fig:mrr_slices}
\end{figure}

\begin{figure}[t]
\centering
\includegraphics[width=0.8\columnwidth]{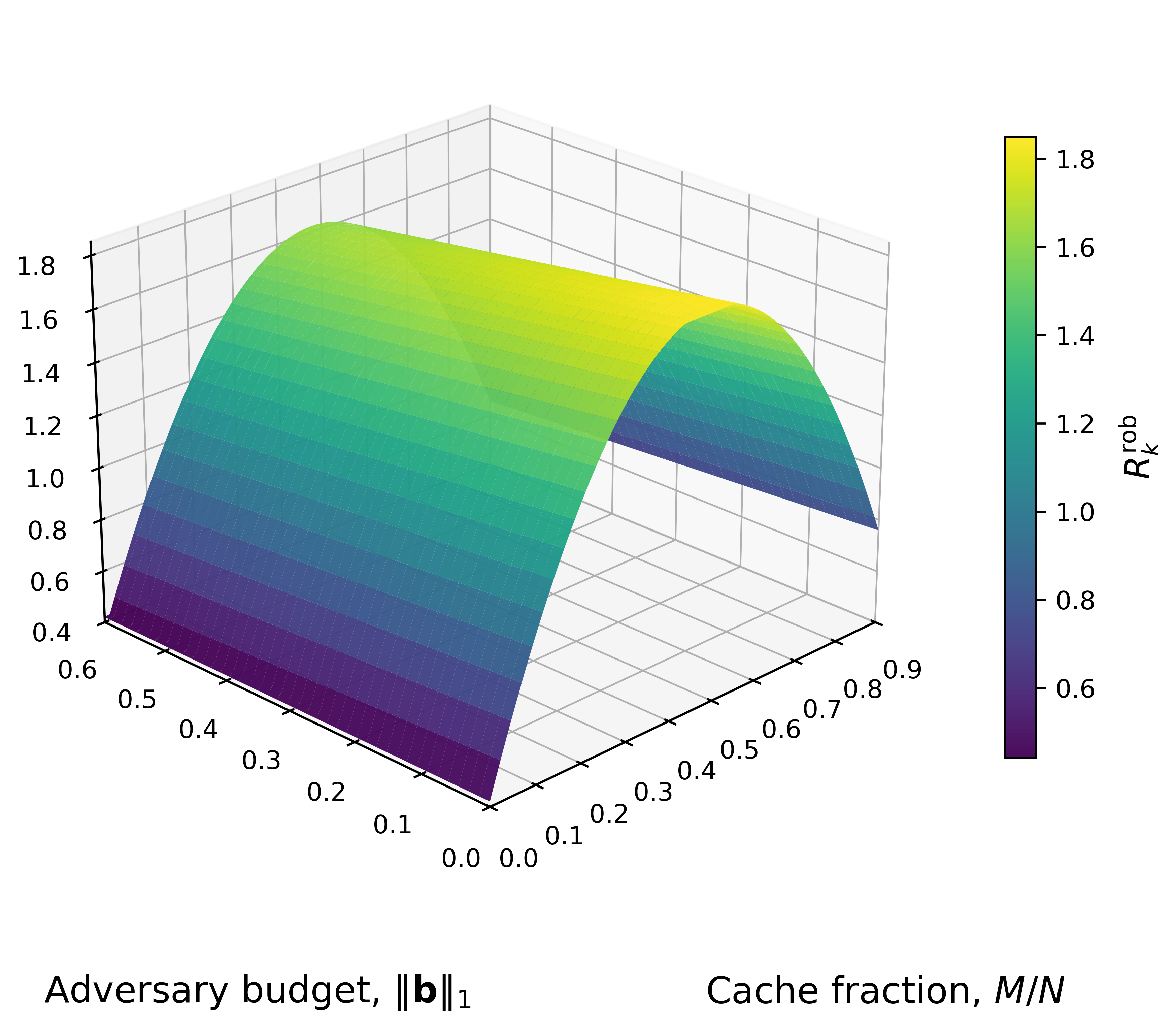}
\caption{MRR surface over $(M/N,\|\mathbf{b}\|_1)$ with breakdown ridge along $f_{\mathcal{B}}=f_{\max}$.}
\label{fig:mrr_surface}
\end{figure}

\begin{figure}[t]
\centering
\includegraphics[width=0.8\columnwidth]{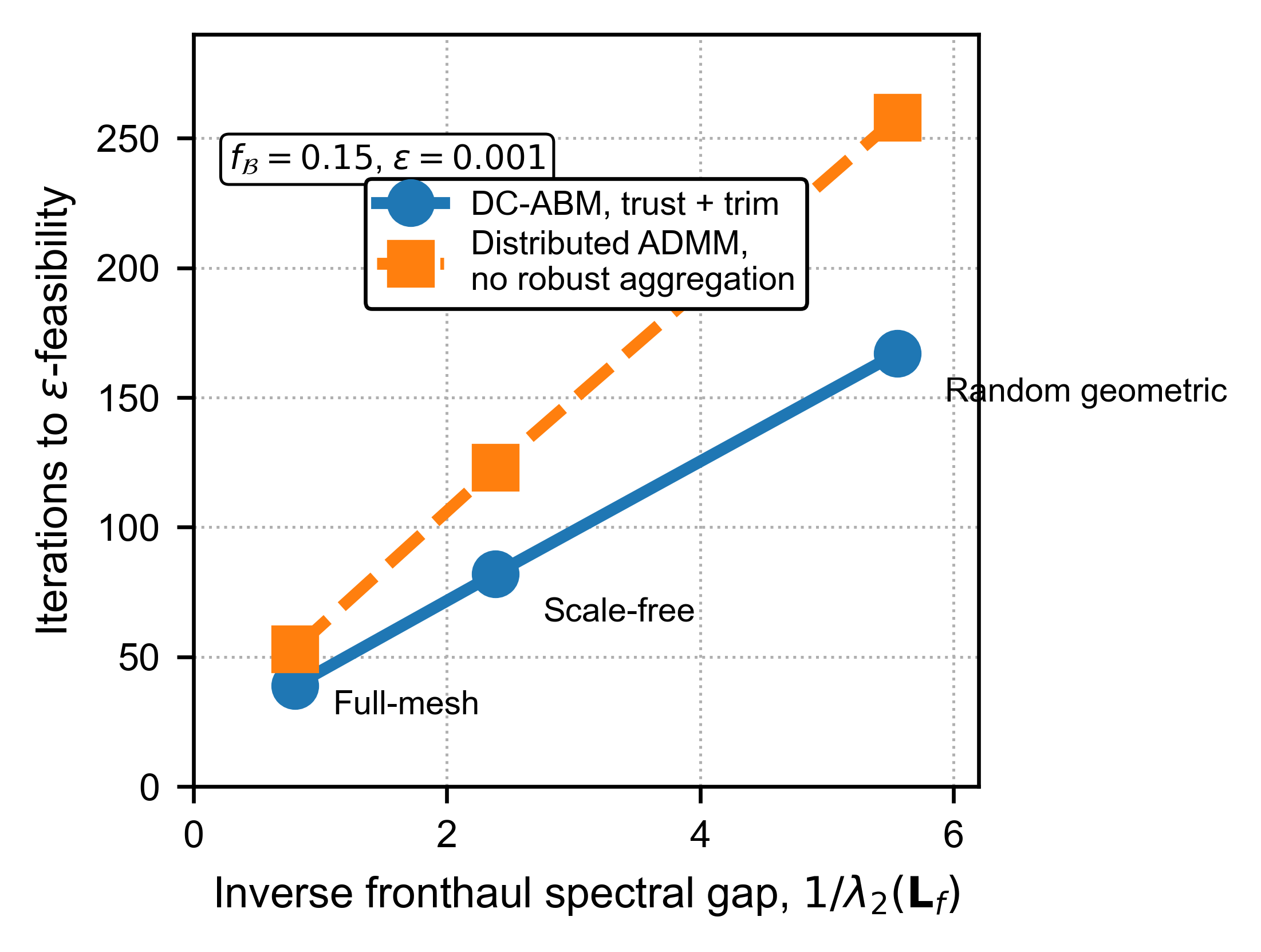}
\caption{Iterations to $\varepsilon$-feasibility vs.\ $\lambda_2(\mathbf{L}_f)$.}
\label{fig:conv_spectral}
\end{figure}

\begin{figure}[t]
\centering
\includegraphics[width=0.8\columnwidth]{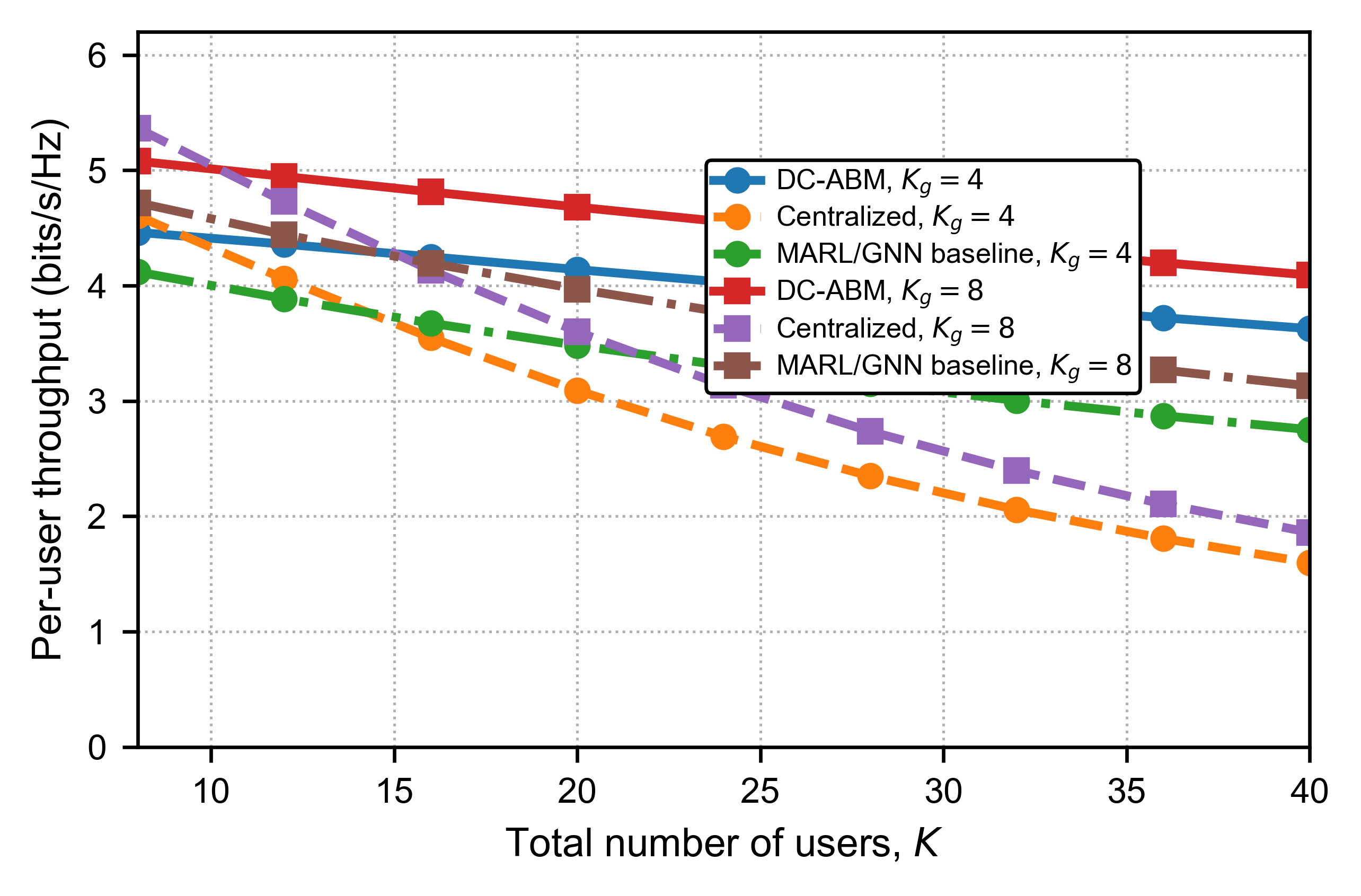}
\caption{Per-user throughput vs.\ $K$ for $K_g\in\{4,8\}$.}
\label{fig:per_user_K}
\end{figure}

\begin{figure}[t]
\centering
\includegraphics[width=0.8\columnwidth]{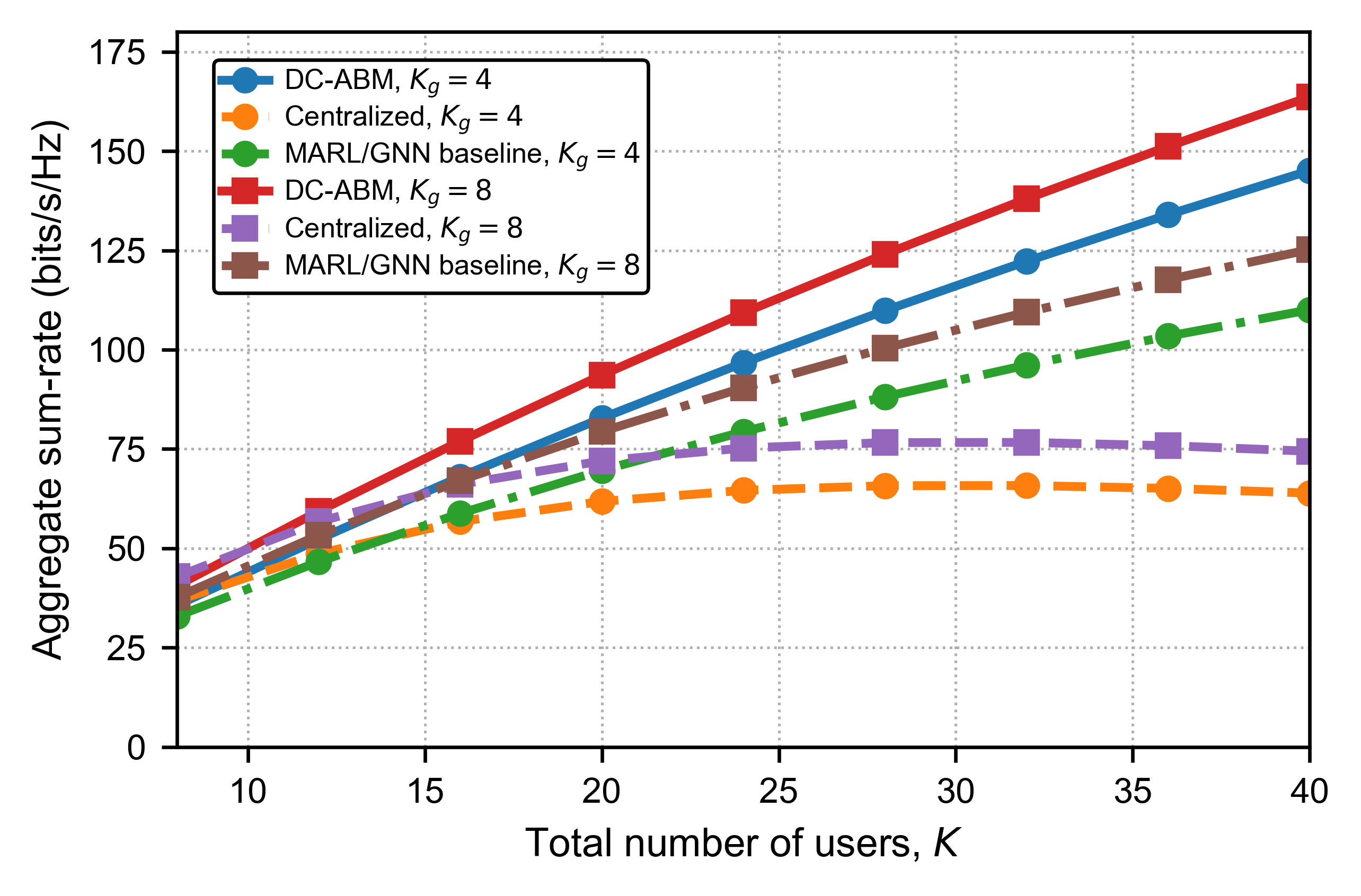}
\caption{Aggregate sum-rate vs.\ $K$ for the same configurations.}
\label{fig:aggregate_K}
\end{figure}

\begin{figure}[t]
\centering
\includegraphics[width=0.8\columnwidth]{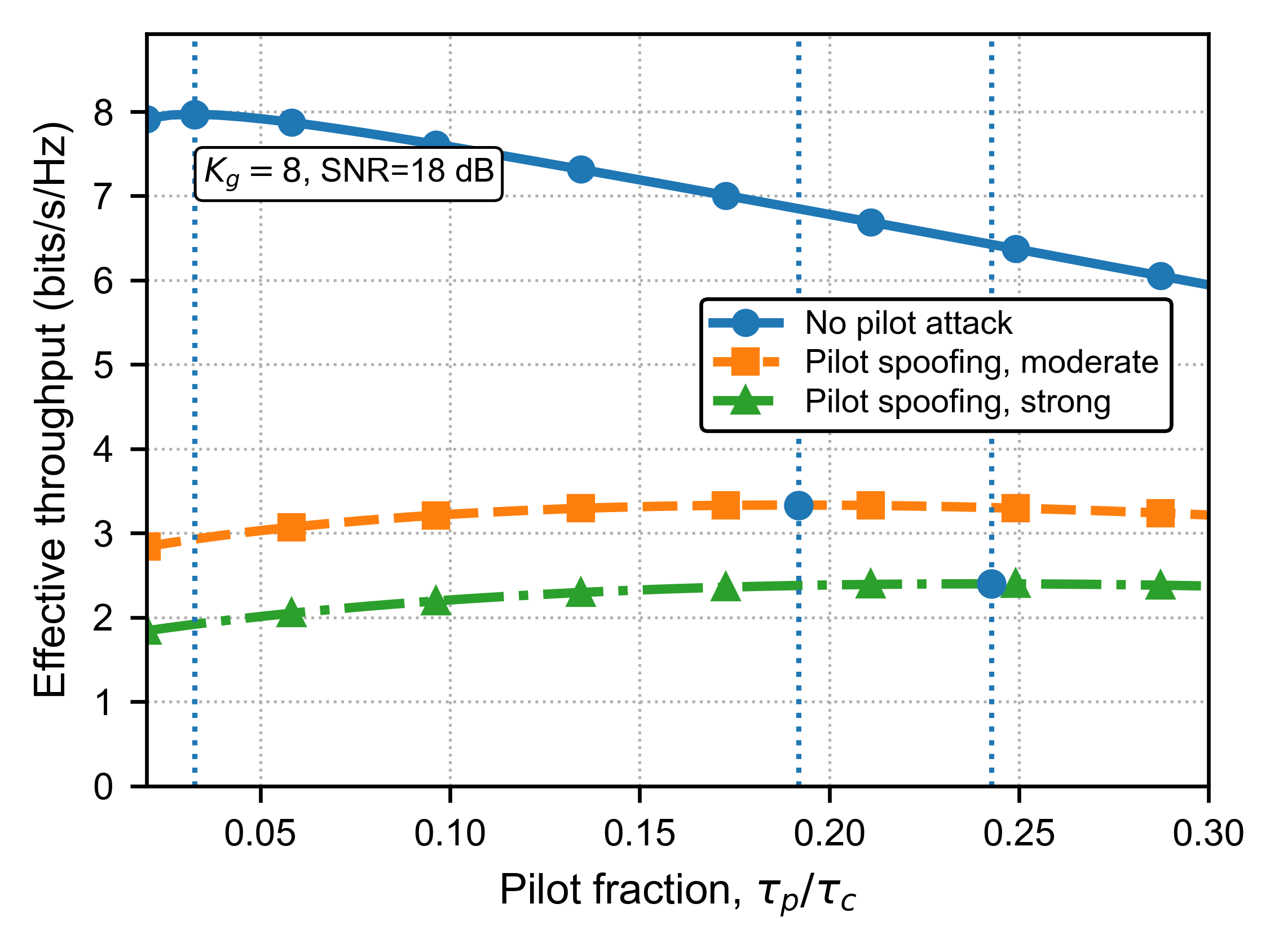}
\caption{Pilot-spoofing sensitivity.}
\label{fig:pilot_spoof}
\end{figure}

Figs.~\ref{fig:mrr_slices} and~\ref{fig:mrr_surface} explore the joint $(M/N,\|\mathbf{b}\|_1)$ trade-off. The slice view shows the achievability bound retains the MN shape multiplied by $\Phi_k(\mathbf{b})$; the surface view exhibits smooth monotone shape with a breakdown ridge along $f_{\mathcal{B}}=f_{\max}$. Fig.~\ref{fig:conv_spectral} confirms $\mathcal{O}(\lambda_2^{-1})$ scaling of convergence iterations. Figs.~\ref{fig:per_user_K} and~\ref{fig:aggregate_K} confirm linear sum-rate scaling in $K$ for two group sizes $K_g\in\{4,8\}$. Fig.~\ref{fig:pilot_spoof} shows pilot-spoofing sensitivity. Relative to the no-spoofing baseline, the effective throughput drops substantially: at a pilot fraction of $\tau_p/\tau_c\approx 0.15$, DC-ABM retains approximately $45\%$ of the baseline throughput under moderate spoofing and approximately $32\%$ under strong spoofing, in agreement with the curves in Fig.~\ref{fig:pilot_spoof}, highlighting the detrimental impact of pilot contamination on channel estimation quality. Nevertheless, the proposed DC-ABM framework maintains non-zero throughput across the entire pilot-fraction range. This robustness is attributable to the EN-partition $\mathrm{DWPR}^{\mathrm{att}}$, which spreads multicast streams across multiple ENs experiencing partially independent channel-estimation errors and contamination realizations. \emph{Physically}, the MRR surface demonstrates that cache memory acts as ``stored adversarial slack'', larger cache fractions increase coded-multicast opportunities and partially compensate for attack-induced performance degradation by providing additional delivery alternatives and redundancy. Furthermore, Figs.~\ref{fig:per_user_K} and~\ref{fig:aggregate_K} reveal the trade-off between cache-induced multiplexing gain and spatial-DoF overhead. Increasing the group size from $K_g=4$ to $K_g=8$ creates more multicast coding opportunities and improves aggregate sum-rate scaling, but it also increases the number of simultaneously served users and the associated spatial-resource requirements. Finally, the pilot-spoofing results highlight the practical role of $\mathrm{DWPR}^{\mathrm{att}}(\mathcal{P}_{\mathrm{EN}})$: rather than allowing pilot contamination to severely degrade a single EN, the distributed architecture averages contamination effects across multiple ENs, thereby reducing sensitivity to localized spoofing attacks.

\vspace{-4mm}
\subsection{Convergence Methodology and Pareto Frontier}

\begin{figure}[t]
\centering
\includegraphics[width=0.8\columnwidth]{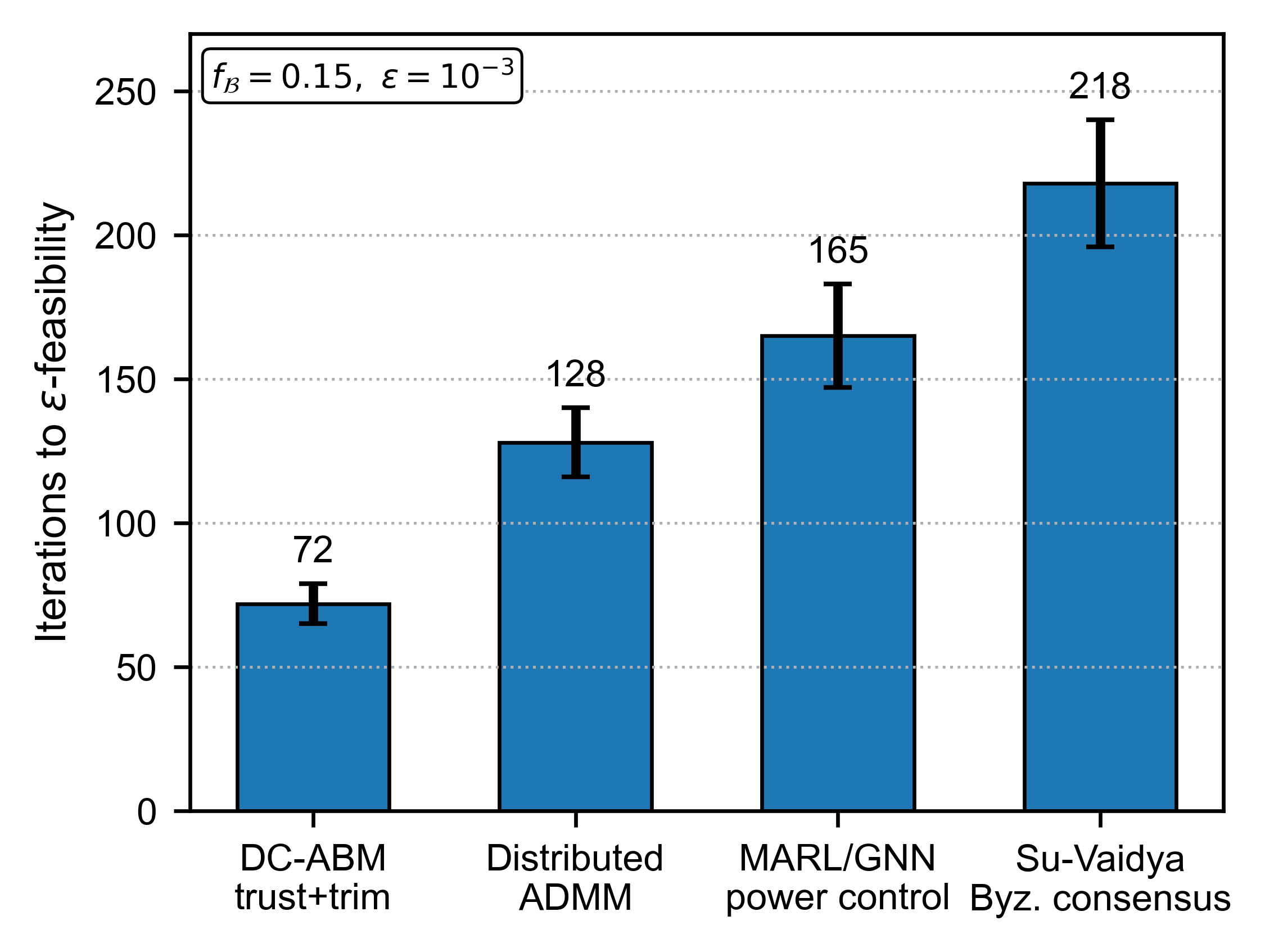}
\caption{Convergence: iterations to feasibility at $f_{\mathcal{B}}=0.15$. DC-ABM is 1.8--3$\times$ faster than baselines.}
\label{fig:conv_methods}
\end{figure}

\begin{figure}[t]
\centering
\includegraphics[width=0.9\columnwidth]{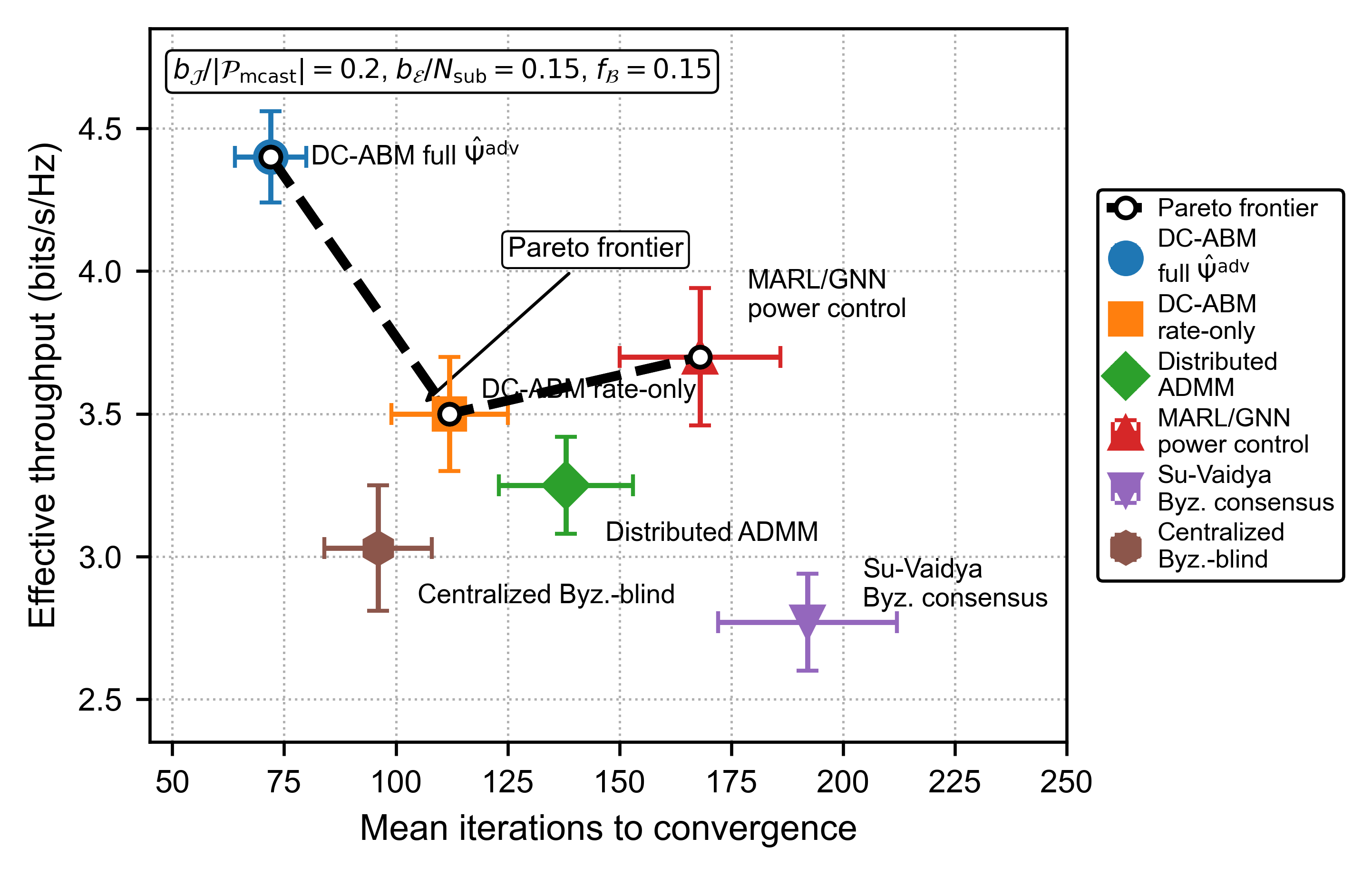}
\caption{Effective throughput versus convergence iterations for the evaluated schemes. DC-ABM achieves the best throughput--convergence trade-off, combining the highest throughput with the fastest convergence.}
\label{fig:pareto}
\end{figure}

Fig.~\ref{fig:conv_methods} reports iterations to $\varepsilon$-feasibility at $f_{\mathcal{B}}=0.15$ for four methods: DC-ABM (72), distributed ADMM (128), MARL/GNN (165), Su--Vaidya (218). DC-ABM is 1.8$\times$ faster than ADMM, 2.3$\times$ faster than MARL/GNN, 3.0$\times$ faster than Su--Vaidya. Fig.~\ref{fig:pareto} shows the effective throughput versus convergence-iteration Pareto frontier at $f_{\mathcal{B}}=0.15$. The proposed DC-ABM occupies the upper-left region of the frontier, simultaneously achieving high throughput and fast convergence. This behavior arises because the structural metrics $\mathrm{DWPR}^{\mathrm{att}}$ and $\mathrm{FSS}^{\mathrm{trust}}$ jointly improve multicast robustness and reduce vulnerability to Byzantine attacks. \emph{Physically}, the iteration-count differences translate directly into delivery-slot latency: for a typical 6G slot duration of $0.1$~ms, $72$ iterations correspond to approximately $7.2$~ms allocation latency for DC-ABM, whereas $220$ iterations for Su--Vaidya correspond to roughly $22$~ms, which may exceed the latency budget of delay-sensitive 6G services.

\vspace{-4mm}
\subsection{State-of-the-Art Comparison}
\label{subsec:sota_3d}

The preceding figures address one or two metrics at a time. We close with two single-plot visualizations that capture the multi-method comparison cleanly.

\begin{figure}[!htbp]
\centering
\includegraphics[width=0.8\columnwidth]{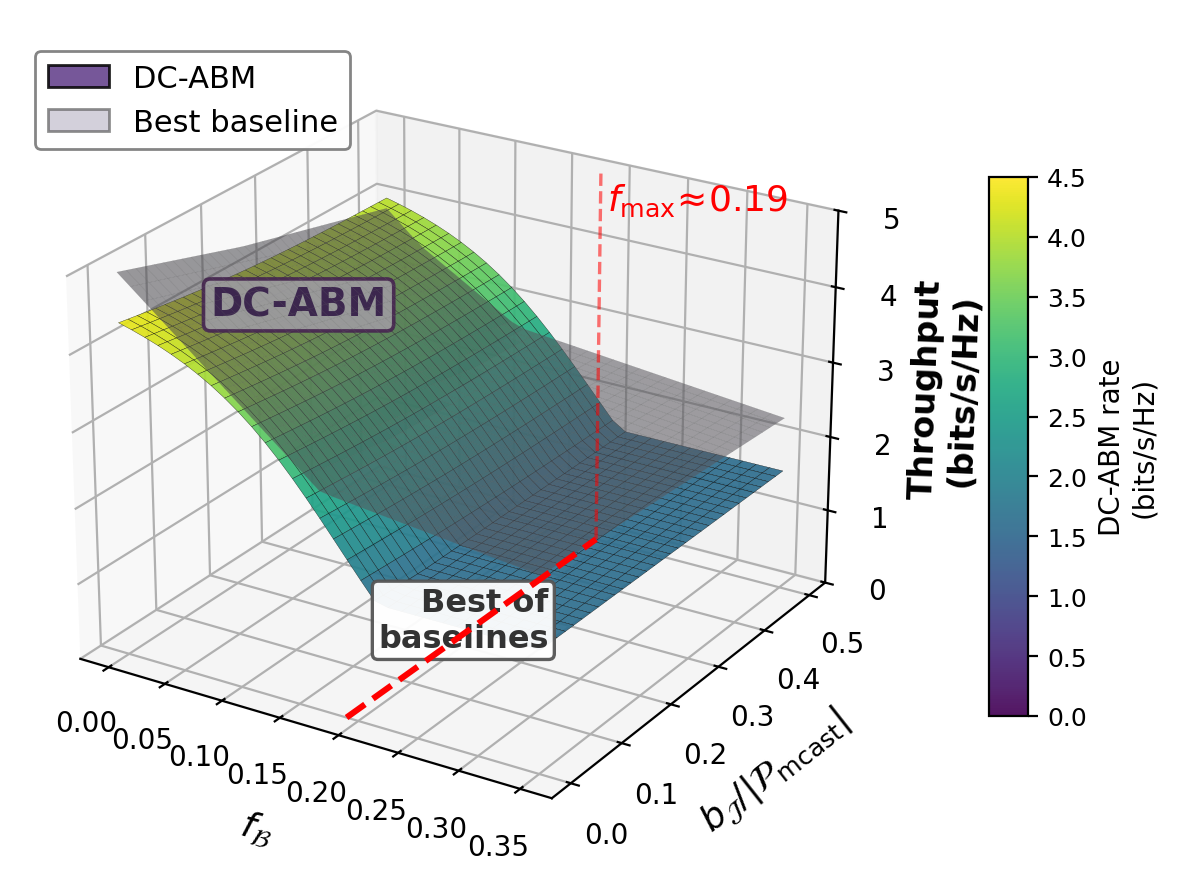}
\caption{Throughput surfaces over the joint adversary plane $(f_{\mathcal{B}},b_{\mathcal{J}}/|\mathcal{P}_{\mathrm{mcast}}|)$: DC-ABM (viridis colormap) versus the pointwise best-of-baselines envelope (gray) computed as $\max(\text{ADMM},\text{MARL/GNN},\text{Su--Vaidya},\text{centralized})$. DC-ABM lies above the envelope throughout the plateau region $f_{\mathcal{B}}<f_{\max}$; the surfaces cross near the red dashed line marking $f_{\max}\approx 0.19$, the breakdown threshold predicted by~\eqref{eq:fmax}, beyond which the gracefully degrading envelope exceeds DC-ABM.}
\label{fig:sota_3d_surfaces}
\end{figure}

Fig.~\ref{fig:sota_3d_surfaces} compares the throughput surface of the proposed DC-ABM framework with the pointwise best-of-baselines envelope across the joint adversary plane $(f_{\mathcal{B}}, b_{\mathcal{J}}/|\mathcal{P}_{\mathrm{mcast}}|)$. The DC-ABM surface exhibits a characteristic ``plateau-then-cliff'' behavior: throughput remains relatively stable and above the baseline envelope over the practically relevant operating region $f_{\mathcal{B}}<f_{\max}$, before decreasing rapidly as $f_{\mathcal{B}}$ approaches the Byzantine tolerance threshold $f_{\max}\approx 0.19$ (red dashed line). In contrast, the baseline envelope degrades gradually without a cliff; consequently, the two surfaces cross in the vicinity of $f_{\max}$, and beyond the breakdown threshold, the envelope dominated there by the centralized scheme, which does not rely on distributed consensus can exceed DC-ABM. This crossover is fully consistent with Theorem~\ref{thm:byz_conv}: once $f_{\mathcal{B}}>f_{\max}$, trust-weighted trimmed aggregation can no longer filter the Byzantine perturbation and no convergence guarantee applies. \emph{Physically}, the separation between the two surfaces in the plateau region reflects the benefit of incorporating attack-aware structural metrics into the resource-allocation process: structural diversity and trust-aware substitution mechanisms absorb moderate adversarial stress with limited throughput loss, whereas beyond $f_{\max}$ the available trustworthy alternatives become insufficient and performance degrades rapidly.

\begin{figure}[!htbp]
\centering
\includegraphics[width=0.8\columnwidth]{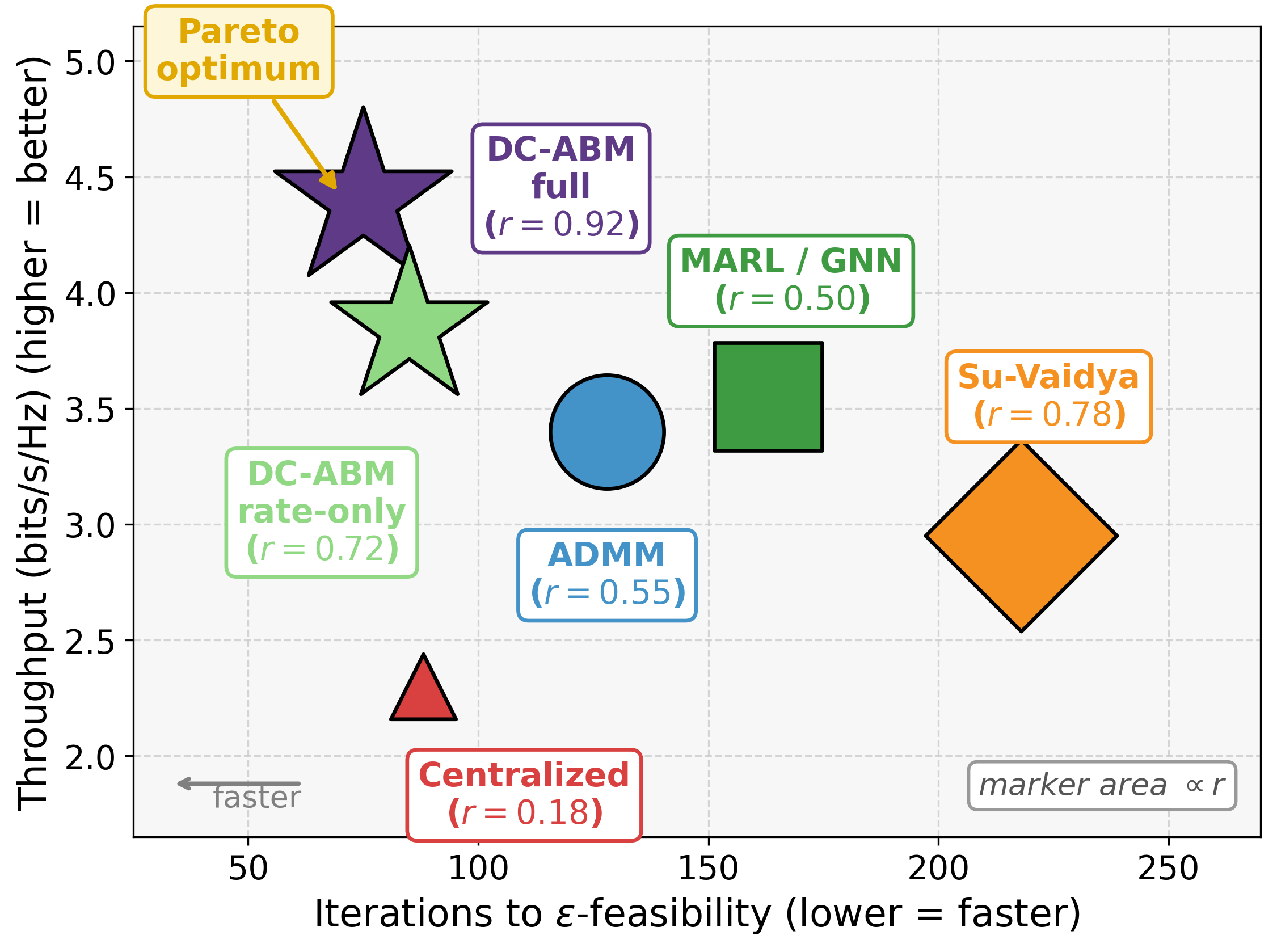}
\caption{Operating-point cloud over three metrics: convergence iterations (horizontal axis, lower is faster), effective throughput (vertical axis, higher is better), and resilience margin $R_k^{\mathrm{rob}}/R_k^{\mathrm{MN}}$ (encoded as marker area, larger is more resilient; numeric value labeled beside each marker). DC-ABM (full $\hat\Psi^{\mathrm{adv}}$, purple star) occupies the Pareto-optimal corner: fewest iterations, highest throughput, and largest resilience margin.}
\label{fig:sota_3d_pareto}
\end{figure}
Fig.~\ref{fig:sota_3d_pareto} compares the considered methods using a Pareto plot in which convergence iterations and throughput define the axes, while the resilience margin $R_k^{\mathrm{rob}}/R_k^{\mathrm{MN}}$ is represented by marker size and annotated numerically. The proposed DC-ABM with full $\hat{\Psi}^{\mathrm{adv}}$ (purple star) occupies the most favorable region, simultaneously achieving the highest throughput, strongest resilience, and lowest iteration count. The DC-ABM rate-only variant (green star) achieves similar throughput but with a noticeably lower resilience margin. Among the baselines, Su--Vaidya provides relatively strong resilience but requires substantially more iterations, whereas ADMM and MARL/GNN achieve moderate throughput and resilience. The centralized scheme is dominated on all metrics and exhibits the smallest resilience margin. The Pareto envelope highlights that no baseline simultaneously matches the throughput, resilience, and convergence performance achieved by DC-ABM.

\section{Conclusion}
\label{sec:conclusion}
We presented a degeneracy-aware framework for resilient resource allocation in CF-CA-MU-MIMO networks under concurrent jamming, cache-aware eavesdropping, and Byzantine cache pollution. The framework rests on three foundations: four canonical vulnerability-mode partitions aligning each attack class with a structurally distinct resource decomposition; three attack-aware structural metrics (DWPR$^{\mathrm{att}}$, FSS$^{\mathrm{trust}}$, $D_k^{\mathrm{rob}}$); and the fully decentralized DC-ABM algorithm with trust-weighted trimmed-mean aggregation and exponential-smoothing trust dynamics. The five theoretical results i.e., a tight top-mass lemma, matching MRR achievability/converse, a multiplicative robust-degeneracy bound with outage corollary, a secrecy--cache--DWPR coupling, and a Byzantine-robust mean-square convergence theorem, characterize the achievable region and reduce exactly to Maddah--Ali--Niesen and Su--Vaidya in the appropriate benign limits. Simulations validate the analytical bounds and demonstrate $1.8\times$ to $3\times$ faster convergence of DC-ABM than distributed ADMM, MARL/GNN-based control, and Su--Vaidya consensus, while maintaining robust throughput performance up to the predicted breakdown threshold $f_{\max}\approx 0.19$. Three extensions are natural: user mobility and dynamic network topology changes that continuously modify trust relationships, user associations, and fronthaul connectivity; active eavesdroppers that probe caches through coordinated request patterns to infer coded-multicast dependencies; and generalization to other shared-resource wireless systems, such as cell-free massive MIMO, O-RAN resource orchestration, distributed edge-computing networks, integrated sensing and communication (ISAC), and semantic communication systems, where partition-indexed redundancy plays a fundamental role in resilience.

\appendices

\section{Proofs}
\label{app:proofs}

\subsection{Proof of Lemma~\ref{lem:topb}}
Order $\boldsymbol{\alpha}$ decreasingly: $\alpha_{(1)}\!\geq\!\cdots\!\geq\!\alpha_{(N)}$. Let $s\!=\!\sum_{i=1}^{b}\alpha_{(i)}$. We show the optimum is two-valued: if two top-$b$ entries differ, averaging them preserves $s$ but decreases $\|\boldsymbol{\alpha}\|^2$, so the saved $\ell_2$-mass can be transferred into the top-$b$ block (maintaining $\ell_1$-sum) to increase $s$. By symmetry the bottom $(N-b)$ block is also constant at optimum: $\alpha_i^+\equiv s/b$ on top-$b$, $\alpha_i^-\equiv(1\!-\!s)/(N\!-\!b)$ otherwise. The $\ell_2$-constraint becomes $s^2/b+(1-s)^2/(N-b)=H$, i.e.\ $(N-b)s^2+b(1-s)^2=bH(N-b)$. Solving this quadratic in $s$ yields~\eqref{eq:topb}; the attainer~\eqref{eq:topb_attainer} follows. $\square$
\vspace{-4mm}
\subsection{Proof of Theorem~\ref{thm:mrr_ach}}
\emph{Step 1 (precoding).} For each $\mathcal{S}\in\mathcal{F}$ set $\mathbf{w}_{\mathcal{S},\ell}$ to the regularized zero-forcing direction onto the worst-case intended user's channel, regularization $\sigma^2_n/P_t$. Under $|\mathcal{F}|\leq LN_a$, the resulting family of multicast precoders $\{\mathbf{w}_{\mathcal{S},\ell}\}_{\mathcal{S}\in\mathcal{F}}$ delivers $\gamma_{k,\mathcal{S}}\geq\gamma^{\mathrm{tar}}$ for $k\in\mathcal{S}$ in benign conditions, yielding $R_k^{\mathrm{MN}}(t)$. \emph{Step 2 (jamming loss).} The budget-$b_{\mathcal{J}}$ jammer attacks $b_{\mathcal{J}}$ streams. Lemma~\ref{lem:topb} applied to $\boldsymbol{\alpha}^{\mathcal{P}_{\mathrm{mcast}}}_k$ with $H=1-\mathrm{DWPR}^{\mathrm{att}}_k(\mathcal{P}_{\mathrm{mcast}})$ bounds the captured fractional rate by~\eqref{eq:topb}, giving multiplier $(1-L_{\mathcal{J}})_+$ with $L_{\mathcal{J}}$ as in~\eqref{eq:LJ}. \emph{Step 3 (Byzantine loss).} A fraction $f_{\mathcal{B}}$ corrupts subfiles, propagating through XOR-cancellation; the FSS$^{\mathrm{trust}}_k$ construction recovers a fraction $\mathrm{FSS}^{\mathrm{trust}}_k$ of corruptions through trusted D2D substitution, giving multiplier $(1-f_{\mathcal{B}}(1-\mathrm{FSS}^{\mathrm{trust}}_k))_+$. Composing yields $\Phi_k(\mathbf{b})$, and~\eqref{eq:mrr_ach} follows. Passive Eve does not enter $\Phi_k$. $\square$
\vspace{-4mm}
\subsection{Proof of Theorem~\ref{thm:mrr_conv}}
Use an oracle adversary. Oracle jamming places $\boldsymbol{\alpha}^{\mathcal{P}_{\mathrm{mcast}}}_k$ at the two-valued worst case~\eqref{eq:topb_attainer} concentrated on a $b_{\mathcal{J}}$-subset; Lemma~\ref{lem:topb} gives captured fraction equal to~\eqref{eq:topb}, yielding rate loss exactly $L_{\mathcal{J}}$. Oracle Byzantine selection picks the $\lfloor f_{\mathcal{B}}K\rfloor$ users intersecting $k$'s decoding XORs across the widest stream set, giving worst-case substitution gap $1-\mathrm{FSS}^{\mathrm{trust}}_k$ and rate loss $f_{\mathcal{B}}(1-\mathrm{FSS}^{\mathrm{trust}}_k)$. Composition gives equality in~\eqref{eq:mrr_ach}. $\square$
\vspace{-4mm}
\subsection{Proof of Theorem~\ref{thm:dk_rob}}
Decompose SINR-denominator perturbations across adversary classes.
\emph{Jamming.} Budget-$b_{\mathcal{J}}$ jammer with per-stream power $P_{\mathrm{jam}}$ contributes $P_{\mathrm{jam}}\sum_{\mathcal{S}\in\mathcal{A}_{\mathrm{mcast}}}|\mathbf{g}_k^{\mathsf{H}}\tilde{\mathbf{w}}_{\mathcal{S}}|^2$. By Lemma~\ref{lem:topb} on the EN-partition mode-mass, the worst-case sum is bounded by $P_{\mathrm{jam}}L_{\mathcal{J}}(b_{\mathcal{J}};\mathrm{DWPR}^{\mathrm{att}}_k(\mathcal{P}_{\mathrm{EN}}))$, yielding~\eqref{eq:DeltaJ}.
\emph{Byzantine.} Each Byzantine user contributes a multiplicative $(1-\mathrm{FSS}^{\mathrm{trust}}_k)$ per affected stream; aggregate gives~\eqref{eq:DeltaB}.
\emph{Eavesdropper.} Passive Eve does not transmit; $\Delta^{\mathcal{E}}_k=0$.
Summing in the $D_k^{\mathrm{rob}}$ definition yields~\eqref{eq:dk_rob_bound}; tightness holds when the three perturbations act on disjoint partitions with additive SINR-denominator effects. $\square$
\vspace{-4mm}
\subsection{Proof of Corollary~\ref{cor:outage}}
Union bound on the system-level definition $D_{\mathrm{sys}}^{\mathrm{rob}}=\max_k D_k^{\mathrm{rob}}$, combined with Theorem~\ref{thm:dk_rob} and $\Delta^{\mathcal{E}}_k=0$, gives~\eqref{eq:outage}. $\square$
\vspace{-4mm}
\subsection{Proof of Theorem~\ref{thm:secrecy}}
Decompose $I(W_{d_k};Y_e^\infty)$ into two paths.
\emph{Direct path.} Subfiles $W^{(g)}_{d_k,\mathcal{T}}\in\mathcal{C}_e$ directly contribute mutual information; the worst-case mass over $|\mathcal{C}_e|=b_{\mathcal{E}}$ subsets is $b_{\mathcal{E}}/\binom{K_g-1}{t}$ relative to $H(W_{d_k})$.
\emph{XOR coupling.} Each observed $c_{\mathcal{S}}$ gives a linear constraint on $|\mathcal{S}|=t+1$ subfiles. If Eve knows $b_{\mathcal{E}}'$ of the $t$ others in $\mathcal{S}\setminus\{k\}$, she partially decodes the target. The residual uncertainty depends on cache-decorrelation via $\boldsymbol{\alpha}^{\mathcal{P}_{\mathrm{sub}}}_k$; Lemma~\ref{lem:topb} with $H=1-\mathrm{DWPR}^{\mathrm{att}}_k(\mathcal{P}_{\mathrm{sub}})$ bounds the coupling mass by the XOR-coupling leakage term on the RHS of~\eqref{eq:eve_info}.
\emph{D2D refinement.} Subfiles delivered through trusted D2D do not enter $Y_e$; the RHS is multiplied by $(1-\rho_k)$. Summing the two paths gives~\eqref{eq:eve_info}. $\square$
\vspace{-4mm}
\subsection{Proof of Theorem~\ref{thm:byz_conv}}

\emph{Step 1 (trim tolerance).} Under (iii), $\beta>f_{\mathcal{B}}$ excises Byzantine reports outside the central $1-2\beta$ window. By Hoeffding's inequality, residual Byzantine influence per iteration is bounded by $b_{\mathrm{trim}}\propto\sigma_\nabla$ with probability $1-\mathcal{O}(e^{-c\beta^2|\mathcal{N}|})$. \emph{Step 2 (trust decay).} Byzantines passing the trim still fail the honest-majority consistency vote~\eqref{eq:consistency}; under Assumption~\ref{assn:honest_majority}, $\tau_j(t)$ decays multiplicatively as $(1-\eta_\tau)^t$, and contribution to~\eqref{eq:trimmed_mean} vanishes exponentially:
$\varepsilon_{\mathcal{B}}(t)\leq b_{\mathrm{trim}}(1-\eta_\tau)^t$,
whose series $\sum_t \varepsilon_{\mathcal{B}}(t)$ is summable. \emph{Step 3 (consensus contraction).} Under (i)--(ii), the primal-dual update takes the form
$\mathbf{P}(t+1)
=
\mathbf{P}(t)
-
\eta_t\mathbf{F}(\mathbf{P}(t))
+
\eta_t\boldsymbol{\xi}(t)$,
with $\kappa$-strongly-monotone $\mathbf{F}$, and Laplacian-driven aggregation contracts at rate $\rho=1-c\lambda_2(\mathbf{L}_f)\eta_t$. \emph{Step 4 (combining).} Standard stochastic-approximation analysis under Steps 1--3 and (iv)--(v) gives the mean-square recursion that yields~\eqref{eq:byz_conv_rate} after substituting structural-metric constants from the trust-decay rate and FSS-conditioned trimmed-mean variance. The breakdown threshold~\eqref{eq:fmax} arises from requiring the denominator variance term to remain positive, which constrains
$
f_{\mathcal{B}}
<
\frac{
\lambda_2(\mathbf{L}_f)
}{
c_3\bigl(1+c_4(1-\overline{\mathrm{FSS}^{\mathrm{trust}}})\bigr)
}$,
giving $c_1=1/c_3$. $\square$
\bibliographystyle{IEEEtran}
\bibliography{TCOM}

\end{document}